\documentclass[prb,aps,twocolumn,amsmath,amssymb,floatfix,
superscriptaddress]{revtex4-2}
\usepackage[dvips]{graphics}
\usepackage[justification=raggedright]{caption}
\usepackage{subcaption}
\usepackage{physics}
\usepackage{color}
\usepackage{float}
\usepackage{natbib}
\usepackage[dvipsnames]{xcolor}
\setcitestyle{numbers}
\setcitestyle{square}
\definecolor{dred}{rgb}{0,0,0.6}
\usepackage{graphicx}    
\usepackage{bm}  
\usepackage{amssymb}  
\usepackage{amsmath}
\usepackage{braket}
\usepackage[mathscr]{euscript}
\usepackage[colorlinks=true, allcolors=blue]{hyperref}
\begin{document}
\title{Higher order topology in a band deformed Haldane model}

\author{Srijata Lahiri and Saurabh Basu \\ \textit{Department of Physics, Indian Institute of Technology Guwahati-Guwahati, 781039 Assam, India}}

\date{\today}

\begin{abstract}
Haldane model is a celebrated tight binding toy model in a 2D honeycomb lattice that exhibits quantized Hall conductance in the absence of an external magnetic field. In our work, we deform the bands of the Haldane model smoothly by varying one of its three nearest neighbour hopping amplitudes ($t_1$), while keeping the other two ($t$) fixed. This breaks the $C_3$ symmetry of the Hamiltonian, while the $M_x*T$ symmetry is preserved. The symmetry breaking causes the Dirac cones to shift from the \textbf{K} and the \textbf{K$'$} points in the Brillouin zone (BZ) to an intermediate \textbf{M} point. This is evident from the Berry curvature plots which show a similar shift in the corresponding values as a function of $\frac{t_1}{t}$. We observe two different topological phases, one being a topological insulator (TI) phase and the other is a higher order topological insulator (HOTI). The Chern number ($C$) remains perfectly quantized at a value of $C=1$ for the TI phase and goes to zero in the HOTI phase. Furthermore the evolution of the Wannier charge center (WCC) as the band is deformed shows a jump in the TI phase indicating a non-trivial bulk. We also study the HOTI phase and diagonalize the real space Hamiltonian on a rhombic supercell to show the presence of in-gap zero energy corner modes. The polarization of the system, namely $p_x$ and $p_y$, are evaluated, along the $x$ and the $y$ directions respectively. We see that both $p_x$ and $p_y$ are quantized in the HOTI phase owing to the presence of the inversion symmetry of the system.
\end{abstract}
\maketitle
\section{\label{sec:level1}Introduction}
The field of topologicals insulators (TI) deals with systems that have a gapped bulk in $d$ dimensions but show topological states in the $d-1$ dimensional boundary \cite{Murakami_2011,RevModPhys.82.3045, PhysRevLett.118.076803, Anomalous_QHE}. Higher order topological insulators (HOTI) are an extension to topological insulators, where robust non-trivial boundary states are found in dimensions less than $d-1$ for a bulk that is $d$ dimensional \cite{PhysRevResearch.3.L042044,PhysRevLett.123.256402,PhysRevB.96.245115,PhysRevLett.119.246402,PhysRevB.97.155305,PhysRevLett.120.026801,Costa2021,Noguchi2021,PhysRevB.101.235403,PhysRevB.106.205111,PhysRevLett.121.075502, PhysRevD.13.3398,PhysRevLett.106.106802, Arnob1, Arnob2, Anomalous_HOTI, Anomalous_floquet, Quasicrystal_HOTI, Graphene_floquet_HOTI}. For instance, second order TIs exhibit the signature of non-trivial topology in $d-2$ dimensions. This gives rise to corner modes for a two-dimensional and hinge modes for a three-dimensional bulk. The conventional definition of bulk-boundary correspondence fails in higher order topology. What we rather find here is a refined bulk-boundary correspondence. Higher order topology has been discussed in the context of a spectrum of systems called topological crystalline insulators which are topological systems protected by crystalline symmetries namely inversion, rotation etc \cite{PhysRevB.100.115403}. The refined bulk boundary correspondence has also been extended to higher order electric multipole insulators where the systems feature localized corner excitations as a result of sharply quantized higher order multipole moments\cite{Wladimir2017, Multipole_moment1}. Material candidates like strained SnTe and surface modified BiSe and BiTe which show a distinct higher order topological phase have also been studied in detail \cite{doi:10.1126/sciadv.aat0346}. Several higher order topological insulators have previously been considered as trivial insulators due to the absence of boundary states. A significant example of this scenario being Bismuth which was considered to be a trivial insulator, but rather shows higher order hinge modes owing to double band inversion in the bulk \cite{Noguchi2021, Bismuth_halide}. In this work, we discuss a similar scenario that shows higher order topology.
\begin{figure}
    \centering
    \includegraphics[width=\columnwidth]{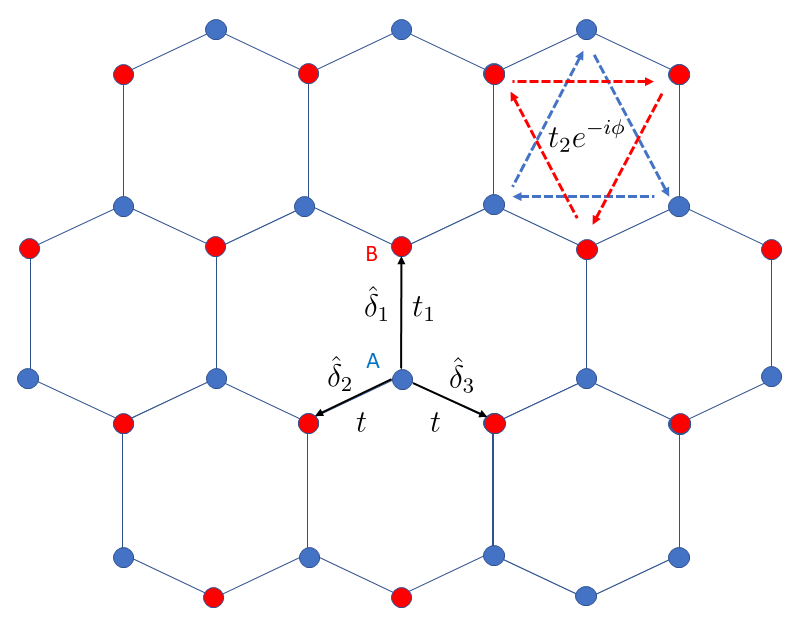}
    \caption{Schematic diagram of the honeycomb lattice. The two sublattices are denoted as $A$ and $B$. The nearest neighbour vector directions are shown as $\hat\delta_1$, $\hat\delta_2$ and $\hat\delta_3$. The hopping along the direction $\hat\delta_1$ is given by $t_1$, while it is given by $t$ in the other two directions. The second neighbour hoppings are given by dashed lines and have an amplitude $t_2$.}
    \label{Fig0}
\end{figure}
\par The Haldane model or Chern insulator is a well known model of a two-dimensional topological insulator built on a hexagonal lattice which exhibits the presence of robust chiral modes at the edges of the system \cite{Haldane}. It was introduced with the motivation to realise non-zero quantized Hall conductance in a honeycomb lattice in the absence of an external magnetic field. The main aim was to gap out the Dirac cones in the bulk Hamiltonian of Graphene by breaking time reversal symmetry via complex next nearest neighbour hopping amplitudes. This opens up non-trivial gaps at the \textbf{K} and \textbf{K$'$} points in the Brillouin zone of the system when the second neighbour hopping (say $t_2$) satisfies the condition $|t_2|\geq \frac{m}{3\sqrt{3}\sin\phi}$. Here $m$ is the Semenoff mass and $\phi$ denotes the phase of the complex hopping amplitude. The system exhibits quantized Hall conductance, which in turn is a result of integral values of the topological invariant, namely the Chern number. The Chern number is a signature of quantized charge transport across the bulk of a system due to adiabatic transformation of the Hamiltonian along a closed loop in the parameter space \cite{TKNN}. This, in turn can be obtained by integrating the Berry phase over the same closed loop. Previous studies on the Haldane model reported a transition from a topological to a trivial phase under a smooth deformation of its bandstructure which is brought about by tuning the amplitude of one of the nearest neighbour (NN) hopping parameter (say $t_1$) while keeping other two NN parameters (say $t$) fixed \cite{Sayan_mondal1}. This deformation breaks the $C_3$ symmetry of the hexagonal lattice. In our work, however, we show that this takes the system from a TI to an HOTI phase. States localised at the corners of a rhombic supercell of the hexagonal lattice have been found when $|\frac{t_1}{t}|$ exceeds a certain critical value. These localized states in dimensions $d-2$ are symbolic of a higher order topological phase. We evaluate the Chern numbers corresponding to both the TI and the HOTI phase. The Chern number exhibited in the HOTI phase is zero, owing to the absence of conducting modes at the edges of the system. Further, we also study the evolution of the Wannier charge center along a closed one dimensional loop in the Brillouin zone both for the TI and the HOTI phase. The Wannier charge center which represents the average position of charge in the unit cell of a crystal lattice bears the same information as is carried by surface energy bands \cite{WCC1, Pos_op}. Furthermore, to characterize our HOTI phase we calculate the polarisation along the $x$ and $y$ directions which are bulk topological invariants obtained as the integral of the Berry connection \cite{MHM, Pyrochlore}.\par The paper is structured as follows. In section II we discuss the tight binding Hamiltonian for the Haldane model and discuss the deformation of the bandstructure as a function of ratio of the two NN hopping amplitudes ($\frac{t_1}{t}$). We define the Chern number as a topological invariant for the TI phase and show that it goes to zero as the system transcends to a higher order topological insulator. We also calculate the Wannier charge center of the system which is equivalent to calculating the Berry phase along one particular direction (say $k_x$) of the 2D BZ and study its evolution over the momentum in the other direction ($k_y$) for both the TI and the HOTI phases. In section III, we plot real space bandstructure of a rhombic supercell of the Haldane model in the HOTI phase and show the presence of zero energy modes that are distinctly separate from the bulk. Further, the probability density of these states are shown which confirm that they are localized at the two corners of the rhombic supercell. Finally, we define bulk polarisation as a topological invariant for the HOTI phase, and show that it is quantized owing to the inversion symmetry of the system.
\section{\label{sec:level2}The Hamiltonian}
\begin{figure}
    \centering
    \includegraphics[height=5cm, width=0.65\columnwidth]{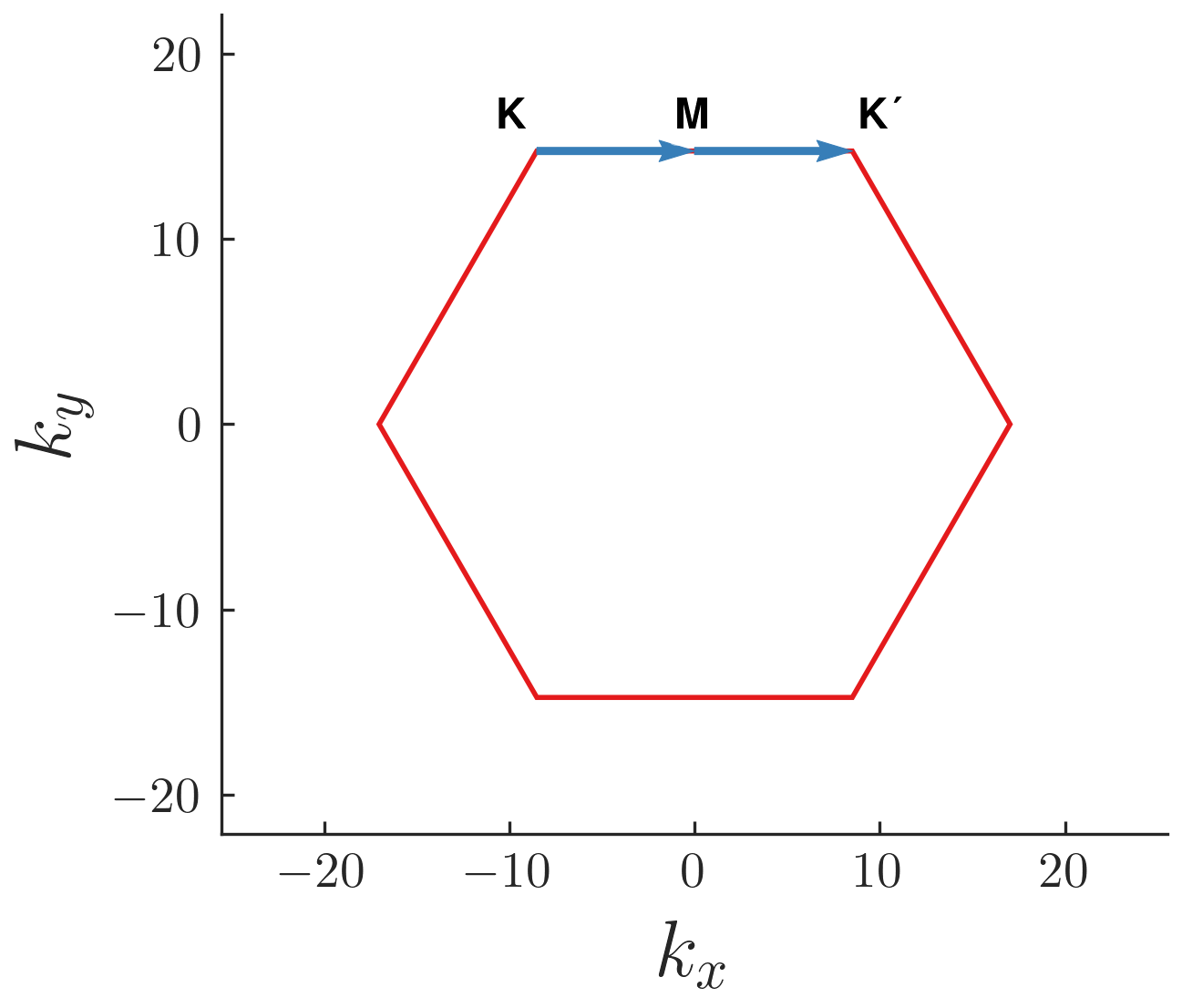}
    \caption{Brillouin zone for the honeycomb lattice with the path (\textbf{K}$\rightarrow$\textbf{M}$\rightarrow$\textbf{K$'$}) shown along which the bandstructure is calculated.}
    \label{Fig1}
\end{figure}
\begin{figure}
\begin{subfigure}[b]{0.49\columnwidth}
         \centering
         \includegraphics[width=\columnwidth]{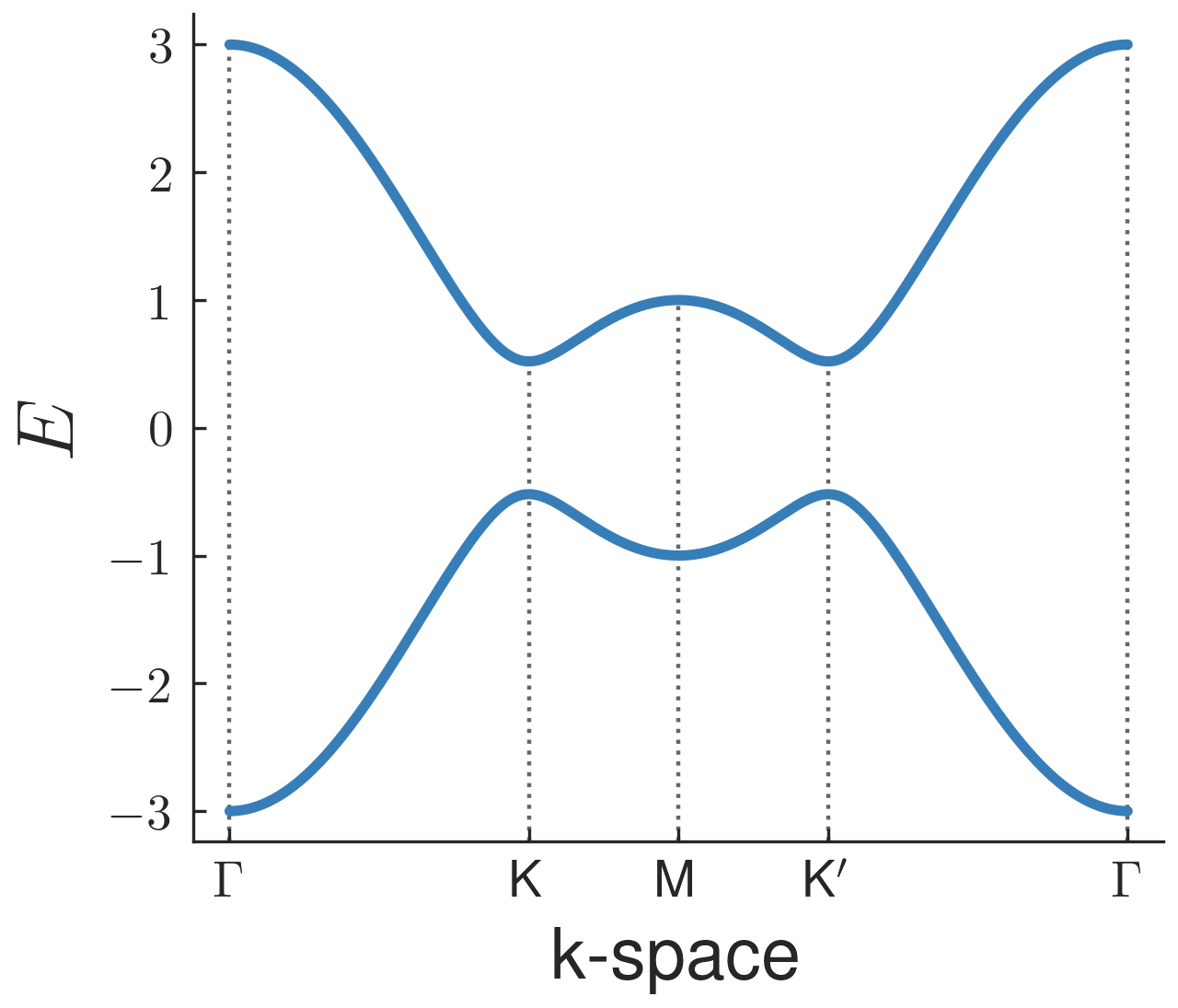}
         \caption{$\frac{t_1}{t}=1$}
         \label{Fig2a}
     \end{subfigure}
\begin{subfigure}[b]{0.49\columnwidth}
         \centering
         \includegraphics[width=\columnwidth]{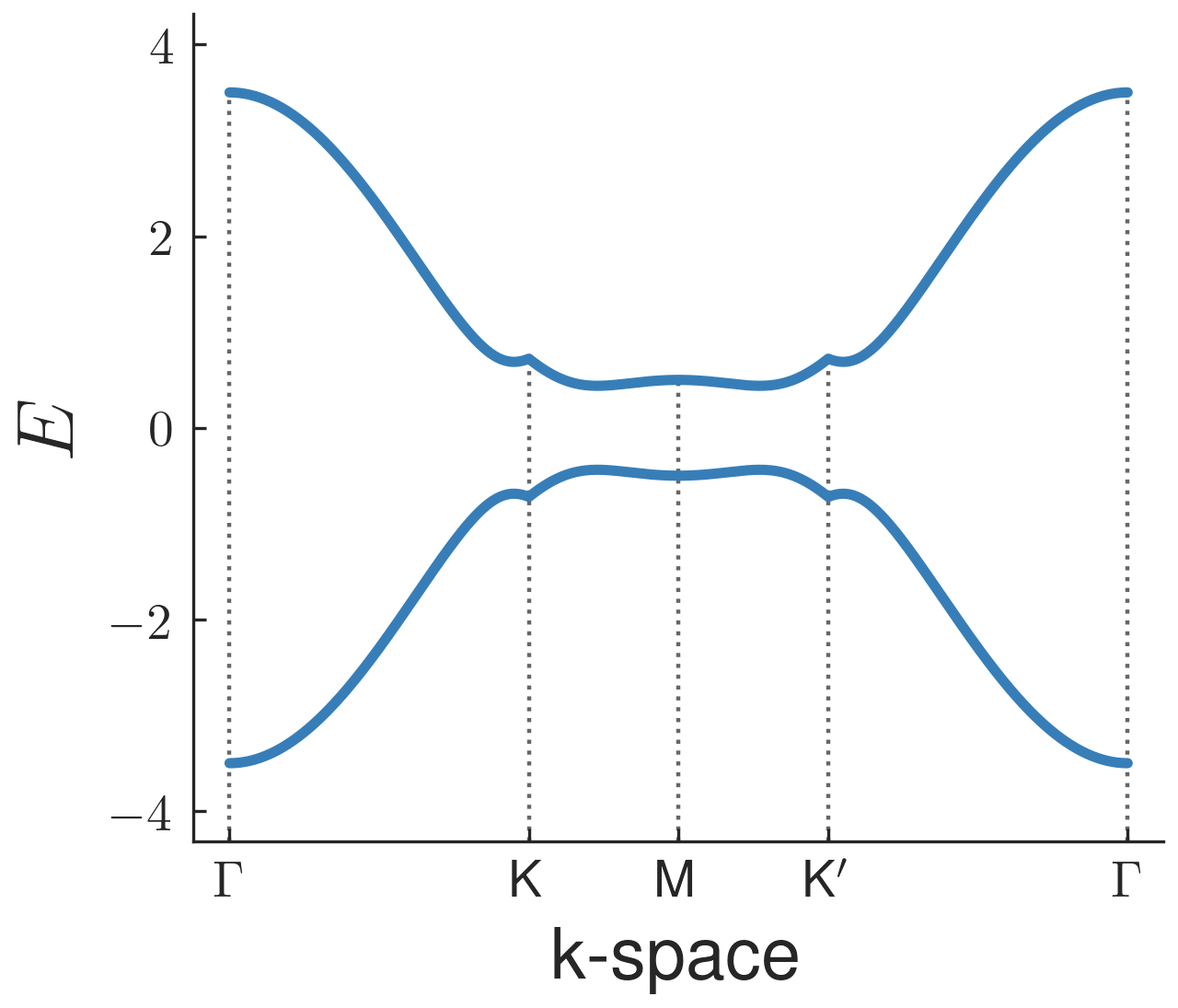}
         \caption{$\frac{t_1}{t}=1.5$}
         \label{Fig2b}
     \end{subfigure}
     \begin{subfigure}[b]{0.49\columnwidth}
         \centering
         \includegraphics[width=\columnwidth]{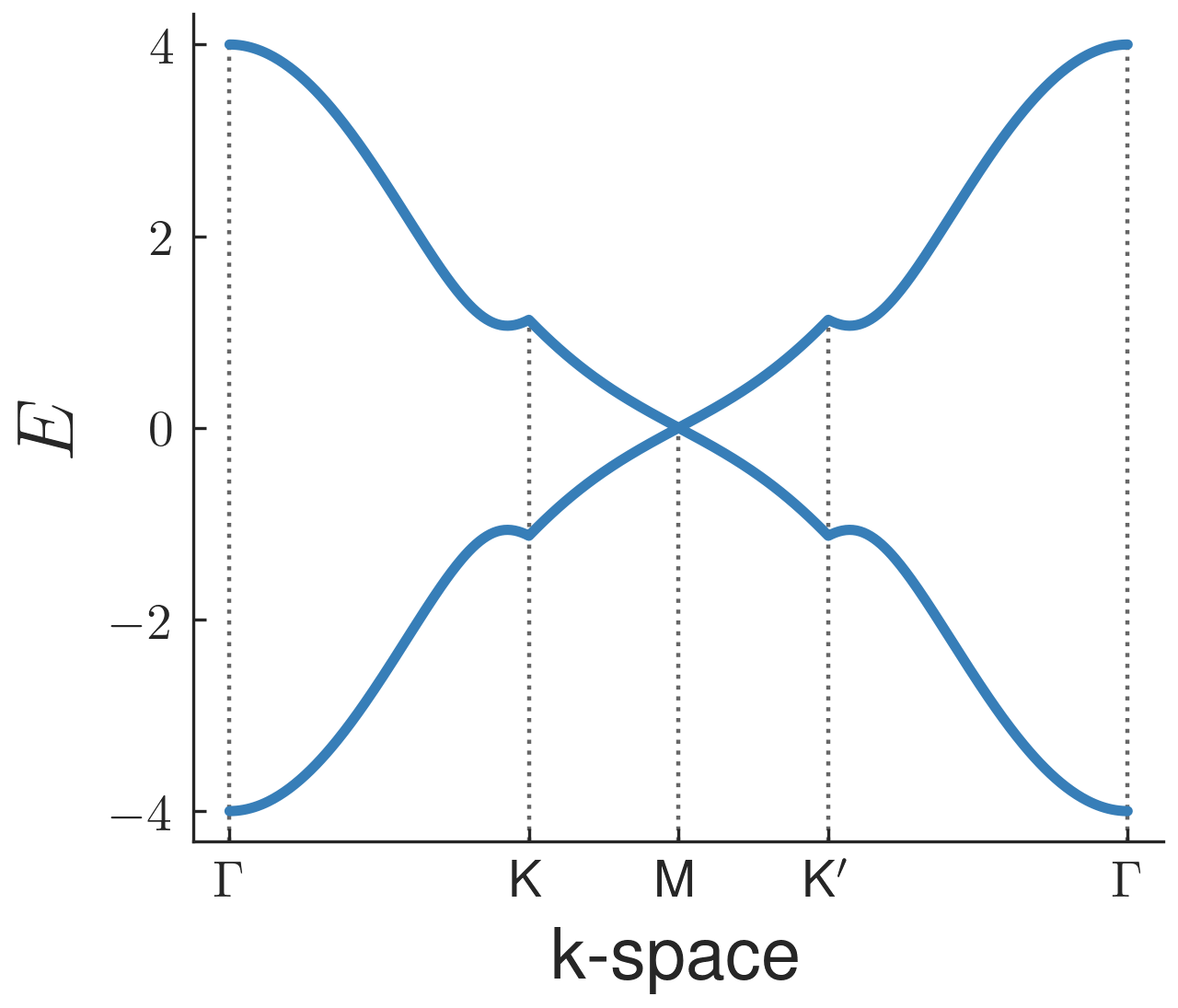}
         \caption{$\frac{t_1}{t}=2.0$}
         \label{Fig2c}
     \end{subfigure}
     \begin{subfigure}[b]{0.49\columnwidth}
         \centering
         \includegraphics[width=\columnwidth]{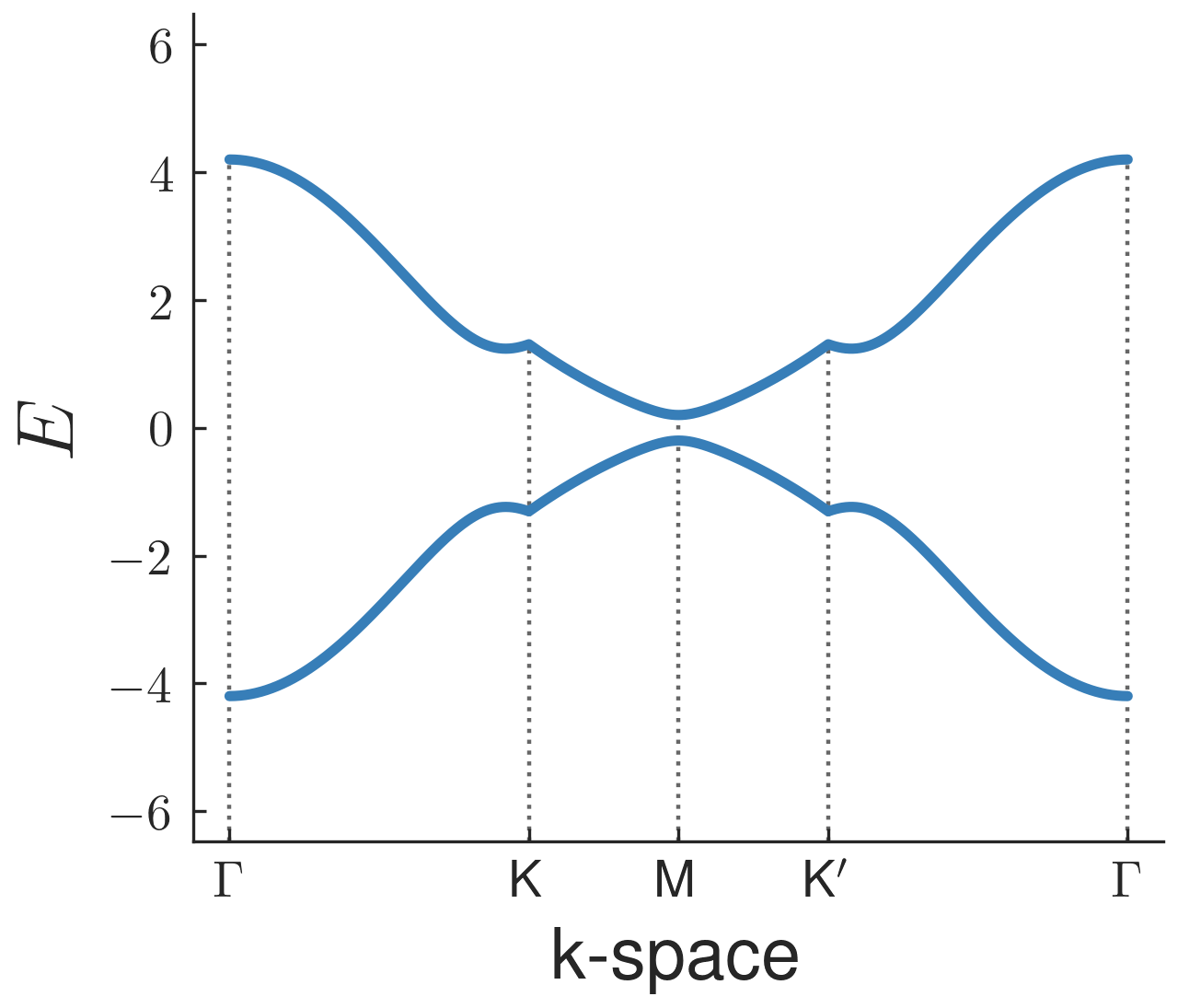}
         \caption{$\frac{t_1}{t}=2.2$}
         \label{Fig2d}
     \end{subfigure}
\caption{The bandstructure of the deformed Haldane model is plotted as a function of $\xi=\frac{t_1}{t}$ along the $\Gamma\rightarrow$\textbf{K}$\rightarrow$\textbf{M}$\rightarrow$\textbf{K$'$}$\rightarrow\Gamma$ path in the BZ of the honeycomb lattice for (a) $\xi=1$, (b) $\xi=1.5$, (c) $\xi=2.0$, (d) $\xi=2.2$. We observe that with increasing $\xi$ the band minima points slowly shift towards the \textbf{M} point of the BZ from the \textbf{K} and \textbf{K$'$} points. At $\xi=2$ the band gap vanishes at the \textbf{M} point indicating a possible topological phase transition. Beyond this point, at $\xi>2$, the band gap reopens. The parameters $t$, $t_2$, $\phi$ and $m$ have been fixed at $1$, $0.1$, $\frac{\pi}{2}$ and $0$. }
 \label{Fig2}
\end{figure}
The Haldane model is defined on a honeycomb lattice as represented in Fig. \ref{Fig0}. The vectors connecting the nearest neighbours are given by $\vec\delta_1=a_0(0,1)$, $\vec\delta_2=a_0(-\frac{\sqrt{3}}{2},-\frac{1}{2})$, $\vec\delta_3=a_0(\frac{\sqrt{3}}{2},-\frac{1}{2})$ where $a_0$ is the length of the nearest neighbour distance. The lattice vectors are given by $\vec{a}_1=\vec{\delta}_1-\vec{\delta}_2$ and $\vec{a}_2=\vec{\delta}_1-\vec{\delta}_3$. In our model, the NN hopping along the direction $\hat \delta_1$ is assumed to be $t_1$, while it is given by $t$ in the directions $\hat\delta_2$ and $\hat\delta_3$. The time reversal symmetry of the system is broken by the next nearest neighbour (NNN) hopping which has a form $t_2e^{i\phi}$ where $\phi$ is the phase associated with the hopping. $\phi$ is positive (negative) for hopping along the anticlockwise (clockwise) direction. For our purpose, we keep $\phi$ fixed at $\frac{\pi}{2}$. The hexagonal lattice has two sublattices denoted by $A$ and $B$ in the figure. The Semenoff mass $m$ is positive(negative) for the sublattice A(B). The NNN vectors, that is, the vectors connecting the nearest neighbour A-A (or B-B) atoms are given by, $\vec\nu_1=\vec\delta_3-\vec\delta_1$, $\vec\nu_2=\vec\delta_2-\vec\delta_3$ and $\vec\nu_3=\vec\delta_1-\vec\delta_2$. The real space Hamiltonian for the Haldane model can be written as,
\begin{align}
    \begin{split}
        H=\sum_{\langle i,j \rangle}t_{ij}c_i^\dagger c_j + \sum_{\langle\langle i,j \rangle\rangle}t_2e^{i\phi}c_i^\dagger c_j + \sum_{i}m_ic_i^\dagger c_i + h.c.
    \end{split}
\end{align}
where $c_i$ ($c_i^\dagger$) represent annihilation (creation) operators at lattice site $i$. We Fourier transform this Hamiltonian to obtain the tight binding Hamiltonian for the deformed Haldane model,
\begin{align}
\begin{split}
H(k)&=\Big{[}t_1\cos\vec{k}.\vec{\delta_1} + \sum_{i=2,3}t\cos{\vec{k}.\vec{\delta_i}}\Big{]}\sigma_x \\&+ \Big{[}t_1\sin\vec{k}.\vec{\delta_1} + \sum_{i=2,3}t\sin{\vec{k}.\vec{\delta_i}}\Big{]}\sigma_y\\& + 2t_2\Big{[}\sin\vec k.(\vec{\delta_3}-\vec{\delta_1})+\sin\vec k.(\vec{\delta_2}-\vec{\delta_3})\\&+\sin\vec k.(\vec{\delta_1}-\vec{\delta_2})\Big{]}\sigma_z+m\sigma_z
\end{split}
\end{align}
Here $\sigma_x$, $\sigma_y$ and $\sigma_z$ denote the Pauli matrices. The Hamiltonian here, has been written in the sublattice basis ($|c_{k_A}\rangle$, $|c_{k_B}\rangle$). It is to be noted that in the original Haldane model, the hopping is uniform along all three nearest neighbour vectors, that is, $t_1=t$. Furthermore, the deformed Haldane model breaks the $C_3$ symmetry which was otherwise present in the original counterpart. It however preserves a product of $M_x$ and $T$, where $M_x$ represents mirror symmetry about $x=0$ and $T$ is the time reversal symmetry operator.\par We first discuss the bandstructure of the deformed Haldane model, as a function of the quantity $\xi=\frac{t_1}{t}$. This system shows two Dirac nodes at the \textbf{K} ($\frac{-2\pi}{3\sqrt3 a_0}, \frac{2\pi}{3a_0}$) and \textbf{K$'$} ($\frac{2\pi}{3\sqrt3 a_0},\frac{2\pi}{3a_0}$) points in the BZ at $\xi=1$ when both $t_2$ and $m$ are zero. The spectral gaps open up at these points when $t_2$ is rendered finite. For our purpose, we keep the value of $t_2$ fixed at $0.1t$ and the phase $\phi$ at $\frac{\pi}{2}$. The Semenoff mass $m$ is also fixed at zero. To aid in our understanding, the BZ of the honeycomb lattice and the path in reciprocal space that has been used for the construction of the bandstructure have been shown in Fig. \ref{Fig1}. As the value of $\xi$ is increased from a value $1$, the gaps in the bandstructure diminish and shift away from the \textbf{K} and \textbf{K$'$} points, as shown in Fig. \ref{Fig2}. As $t_1$ becomes equal to $2t$ ($\xi$=2), the gap closes at the \textbf{M} point $(0, \frac{2\pi}{3a_0})$ in the BZ even when the time reversal symmetry remains broken. On increasing $\xi$ further, the gap reopens. As reported in previous studies, this gap closing was considered to induce a transition of the system from a TI to a trivial phase \cite{Sayan_2}. We however show that beyond $\xi=2$, the system enters into a HOTI phase. \par
\begin{figure}
\begin{subfigure}[b]{0.49\columnwidth}
         \centering
         \includegraphics[width=\columnwidth]{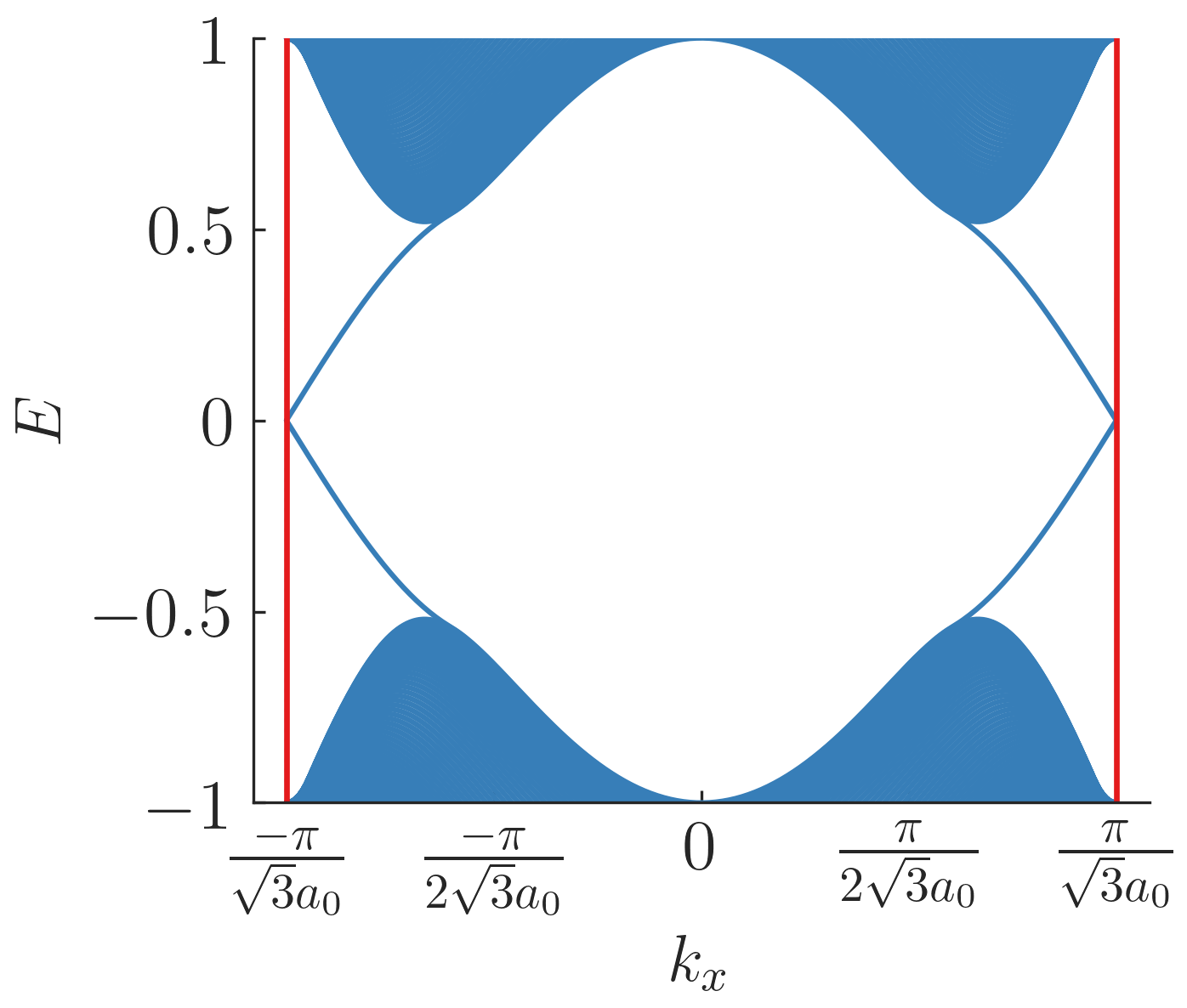}
         \caption{$\xi=1.0$}
         \label{Fig3a}
     \end{subfigure}
\begin{subfigure}[b]{0.49\columnwidth}
         \centering
         \includegraphics[width=\columnwidth]{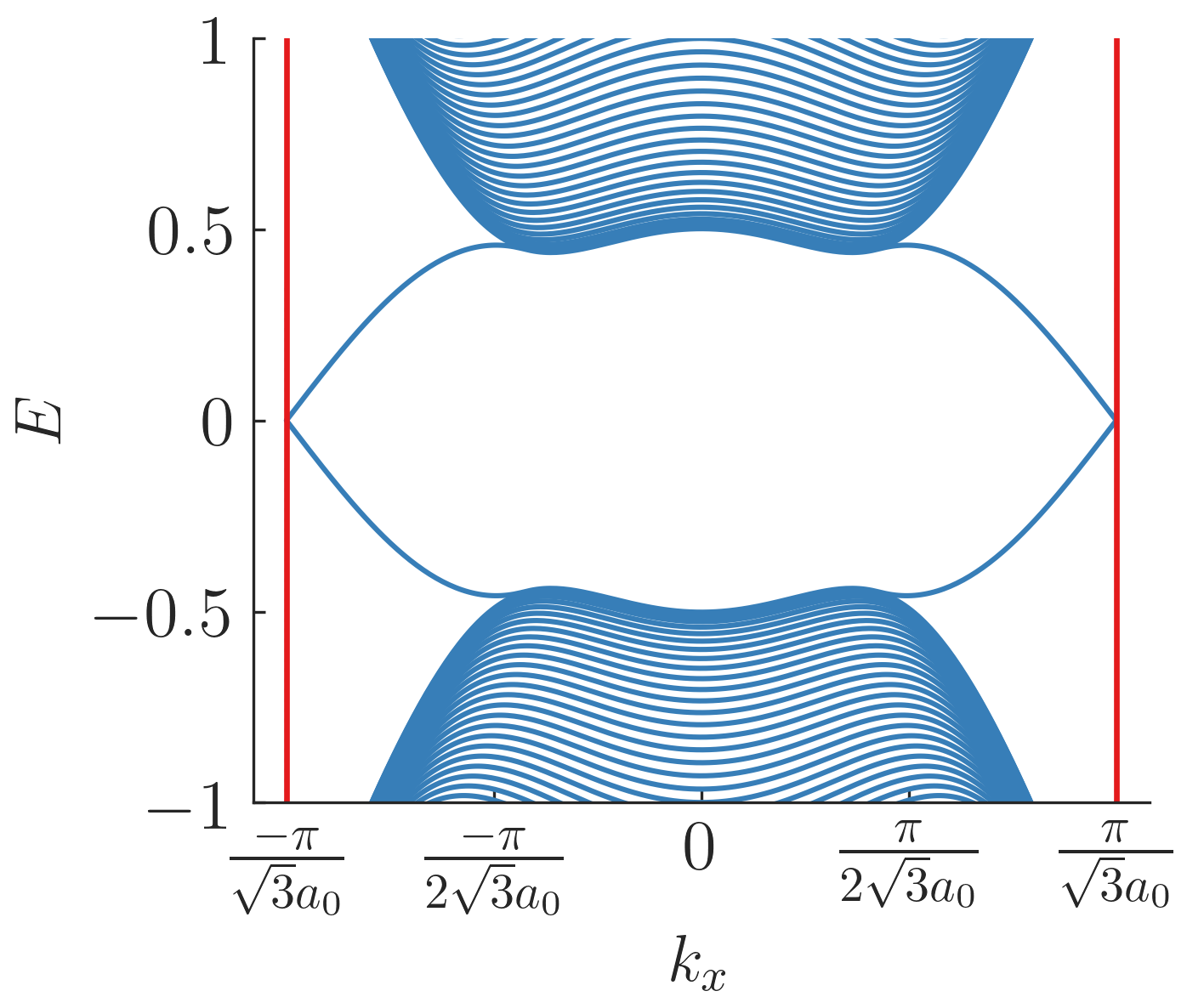}
         \caption{$\xi=1.5$}
         \label{Fig3b}
     \end{subfigure}
     \begin{subfigure}[b]{0.49\columnwidth}
         \centering
         \includegraphics[,width=\columnwidth]{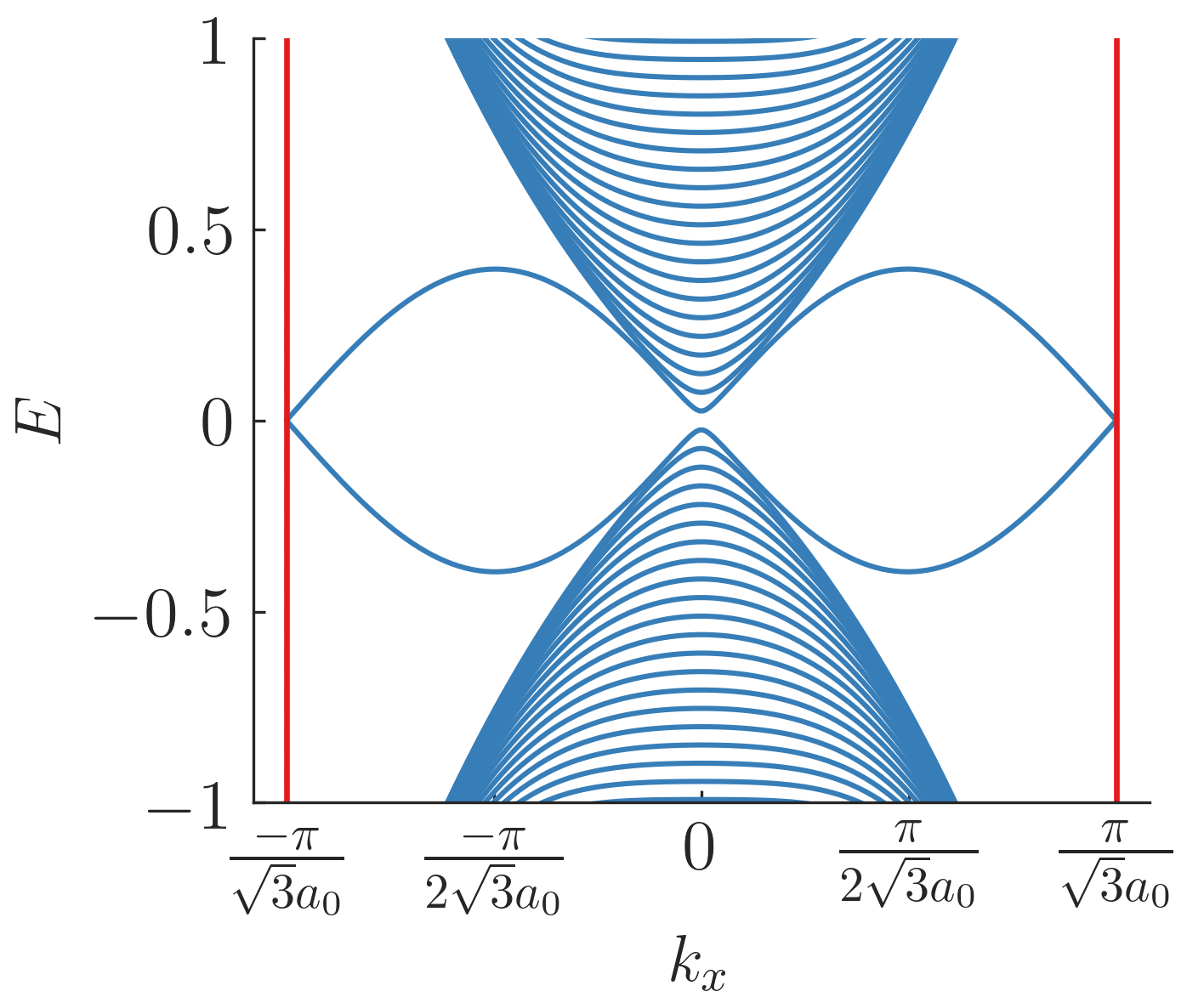}
         \caption{$\xi=2.0$}
         \label{Fig3c}
     \end{subfigure}
     \begin{subfigure}[b]{0.49\columnwidth}
         \centering
         \includegraphics[width=\columnwidth]{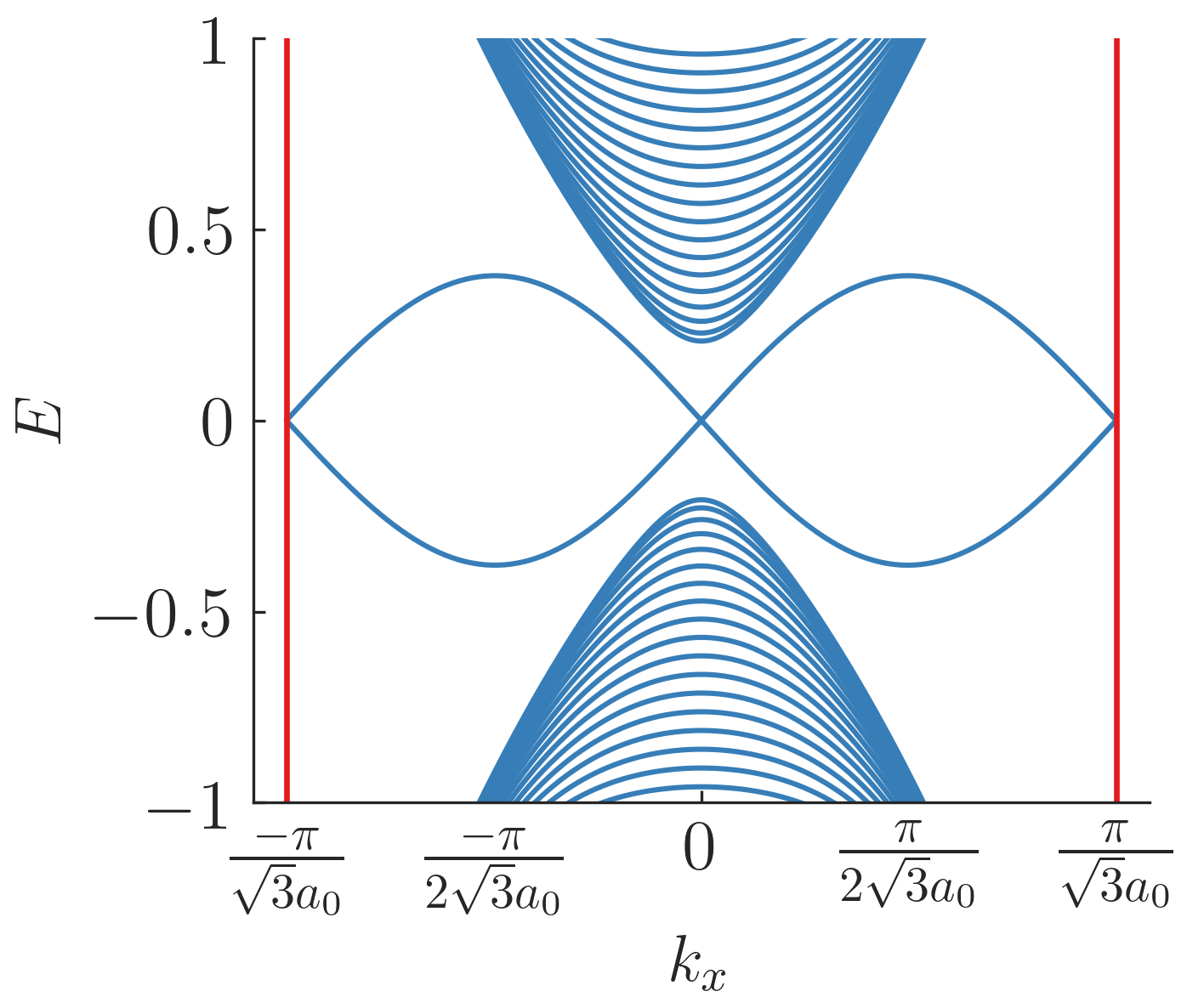}
         \caption{$\xi=2.2$}
         \label{Fig3d}
     \end{subfigure}
\caption{The bandstructure for a zig-zag ribbon-like configuration of the Haldane model is shown. For $\xi<2$ in (a) and (b), we see edge modes crossing the bulk gap. The system is in the TI phase in this region. (c) At $\xi=2$ the bandstructure undergoes a gap closing transition. The closure is not perfect due to the limited number of lattice sites taken along $\hat a_2$. (d) Beyond $\xi=2$ we still see the presence of in-gap modes. However they do not traverse the bulk gap and can be easily removed by an adiabatic deformation of the Hamiltonian. We have kept the values of the parameters $t$, $t_2$, $\phi$ and $m$ fixed at $1$, $0.1$, $\frac{\pi}{2}$ and $0$. The number of unit cells along $\hat a_2$ (see text) is 127.}
 \label{Fig3}
\end{figure}
Next, we study the bandstructure of a zig-zag semi-infinite ribbon-like configuration of the model with periodic boundary condition (PBC) along direction $\hat a_1$ and open boundary condition (OBC) along the direction $\hat a_2$, to show the presence of edge modes connecting the conduction and valence bands in the TI phase (Fig. \ref{Fig3}). The width of the ribbon along the $y$ direction is given by $a_0(\frac{3N}{2}+1)$, where $N$ denotes the number of unit cells in the $y$ direction. We have chosen $N$ to be equal to 127. It can be observed that for $\xi<2$, the system shows the presence of modes that traverse the bulk energy gap (\ref{Fig3a}, \ref{Fig3b}). Exactly at $\xi=2$ (\ref{Fig3c}), the gap closes (the tiny gap seen in the figure at $k_x=0$ will vanish for large longitudinal system dimensions). For $\xi>2$, the bandstructure still shows the presence of in gap modes. However they are completely detached from the bulk and carry no topological significance (\ref{Fig3d}). These modes correspond to counterpropagating edge states that cancel each other at each boundary of the ribbon and gives rise to a trivial insulating phase.\par
To characterize this topological insulator phase, we evaluate the Chern number which is a topological invariant obtained by integrating the Berry connection over a closed loop in the BZ. It is directly related to the conductivity of the edge modes. To calculate the Chern number, we first calculate the Berry curvature of the system. The Berry curvature is given as \cite{Chern_number_fukui, Chern_number2},
\begin{align}
    \begin{split}
       \Omega(\vec k) &=  \curl{\vec A(\vec k)}\\&=i\biggl{[} \biggl <\frac{\partial \psi(\vec k)}{\partial k_x}\biggl|\frac{\partial \psi(\vec k)}{\partial k_y}\biggl> - \biggl <\frac{\partial \psi(\vec k)}{\partial k_y}\biggl|\frac{\partial \psi(\vec k)}{\partial k_x}\biggl>\biggl ]
    \end{split}
\end{align}
\begin{figure}
\begin{subfigure}[b]{0.49\columnwidth}
         \centering
         \includegraphics[width=\columnwidth]{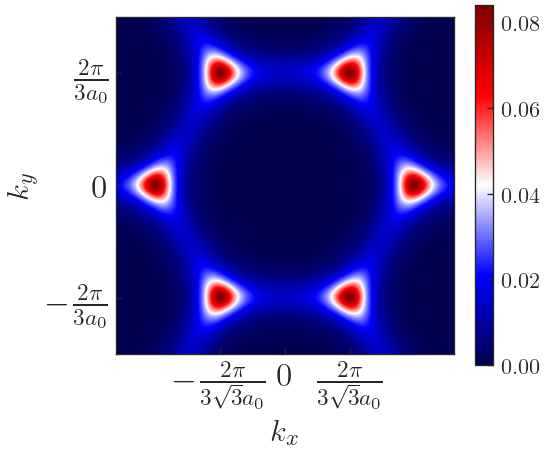}
         \caption{$\frac{t_1}{t}=1.0$}
         \label{Fig4a}
     \end{subfigure}
\begin{subfigure}[b]{0.49\columnwidth}
         \centering
         \includegraphics[width=\columnwidth]{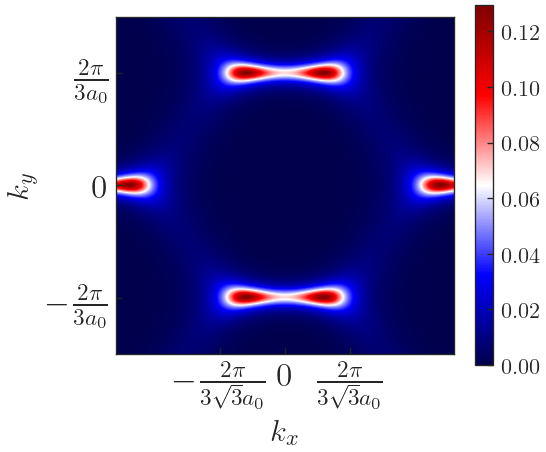}
         \caption{$\frac{t_1}{t}=1.5$}
         \label{Fig4b}
     \end{subfigure}
     \begin{subfigure}[b]{0.49\columnwidth}
         \centering
         \includegraphics[width=\columnwidth]{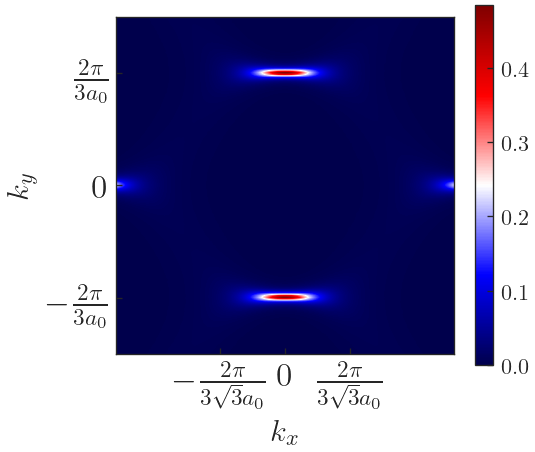}
         \caption{$\frac{t_1}{t}=1.8$}
         \label{Fig4c}
     \end{subfigure}
     \begin{subfigure}[b]{0.49\columnwidth}
         \centering
         \includegraphics[width=\columnwidth]{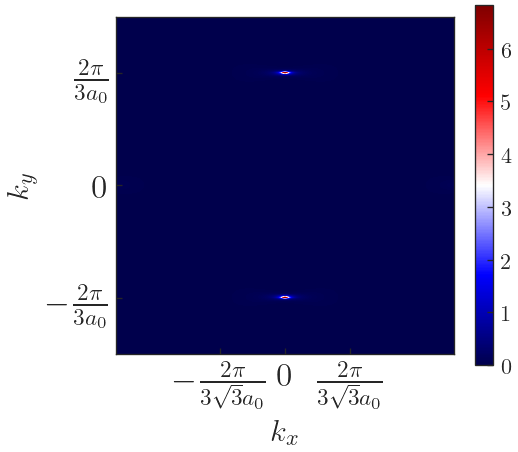}
         \caption{$\frac{t_1}{t}=1.95$}
         \label{Fig4d}
     \end{subfigure}
\caption{The Berry curvature for different values of $\xi=\frac{t_1}{t}$ are shown over the Brillouin zone of honeycomb lattice. For (a) $\xi=1$, the Berry curvature is confined near the \textbf{K} and \textbf{K$'$} points of the BZ. With increasing values of $\xi$ in (b), (c) and (d) the region of high concentrations shift towards the \textbf{M} point. This indicates a shift in the point of minimum band gap with increasing $\xi$.}
 \label{Fig4}
\end{figure}
The Chern number is now calculated using,
\begin{align}
    \begin{split}
        C=\frac{1}{2\pi}\iint_{BZ}\Omega(\vec k).d\vec{S}
    \end{split}
\end{align}
\begin{figure}
    \centering
    \includegraphics[width=\columnwidth]{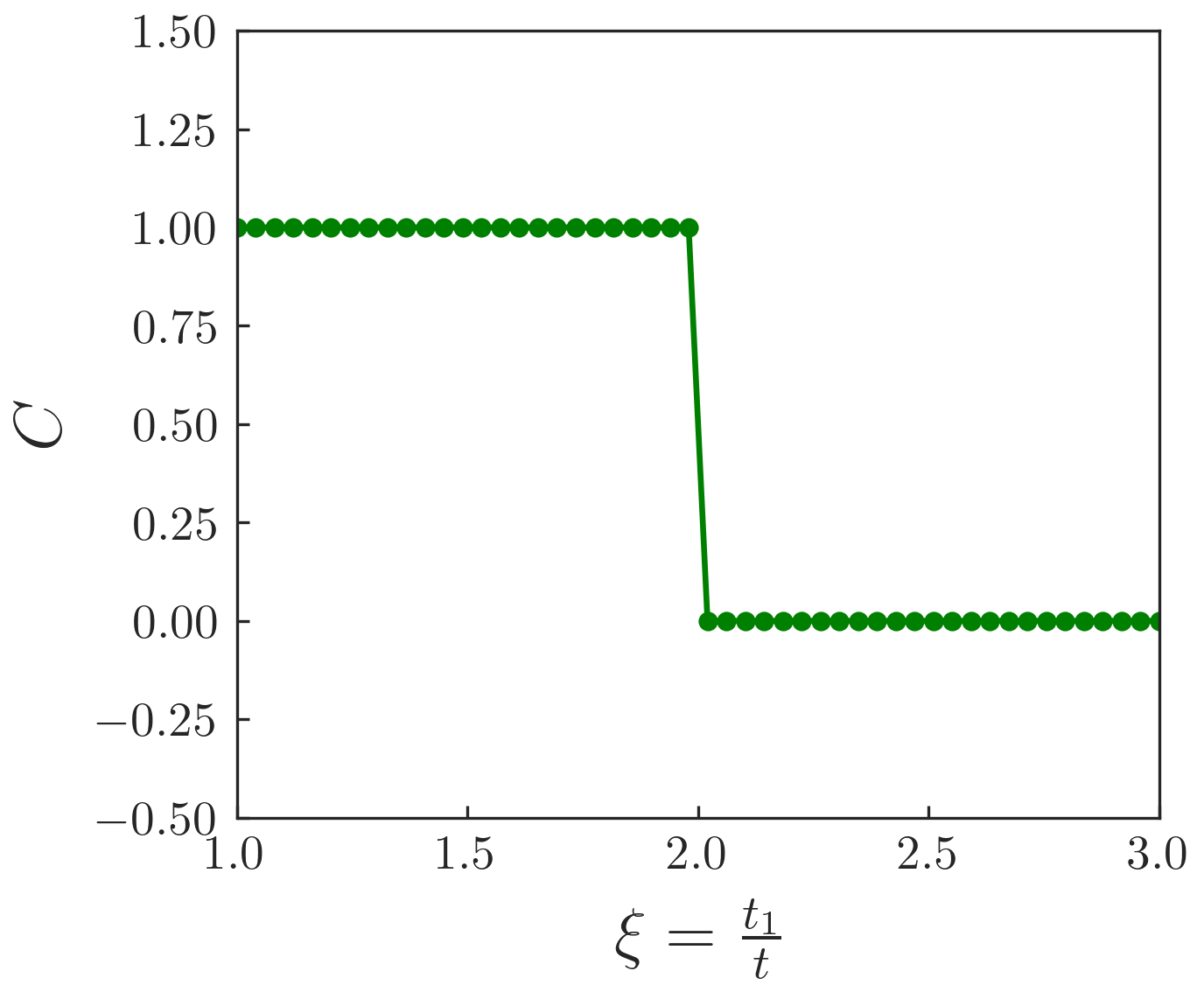}
    \caption{The Chern number is perfectly quantized at $1$ for $\xi<2$, which is the TI phase. For $\xi>2$, the Chern number goes to $0$. This indicates vanishing of the edge modes in the system which is evident from the bandstructure plots (Fig. \ref{Fig3}) as well.}
    \label{Fig5}
\end{figure}

We first plot the Berry curvature of the system over the entire BZ for different values of $\xi$ (Fig. \ref{Fig4}). A non-zero Berry curvature is a direct consequence of the presence of Dirac nodes in the system. Our plot shows that at $\xi=1$, the Berry curvature is concentrated around the \textbf{K} and \textbf{K$'$} points in the BZ (\ref{Fig4a}). As $\xi$ is increased, the Berry curvature slowly shifts towards the $M$ point (\ref{Fig4b}, \ref{Fig4c}, \ref{Fig4d}). This is because with the increase in $\xi$, the $C_3$ symmetry, which kept the Dirac nodes pinned to the \textbf{K} and \textbf{K$'$} points is broken. We integrate this Berry curvature over the BZ to obtain the Chern number $C$. We plot $C$ as a function of $\xi$, as shown in Fig. \ref{Fig5}. It can be clearly seen that $C$ is perfectly quantized at $1$ when $\xi<2$. Furthermore, for $\xi>2$, $C$ goes to $0$.\par
\begin{figure}
\begin{subfigure}[b]{0.49\columnwidth}
         \centering
         \includegraphics[width=\columnwidth]{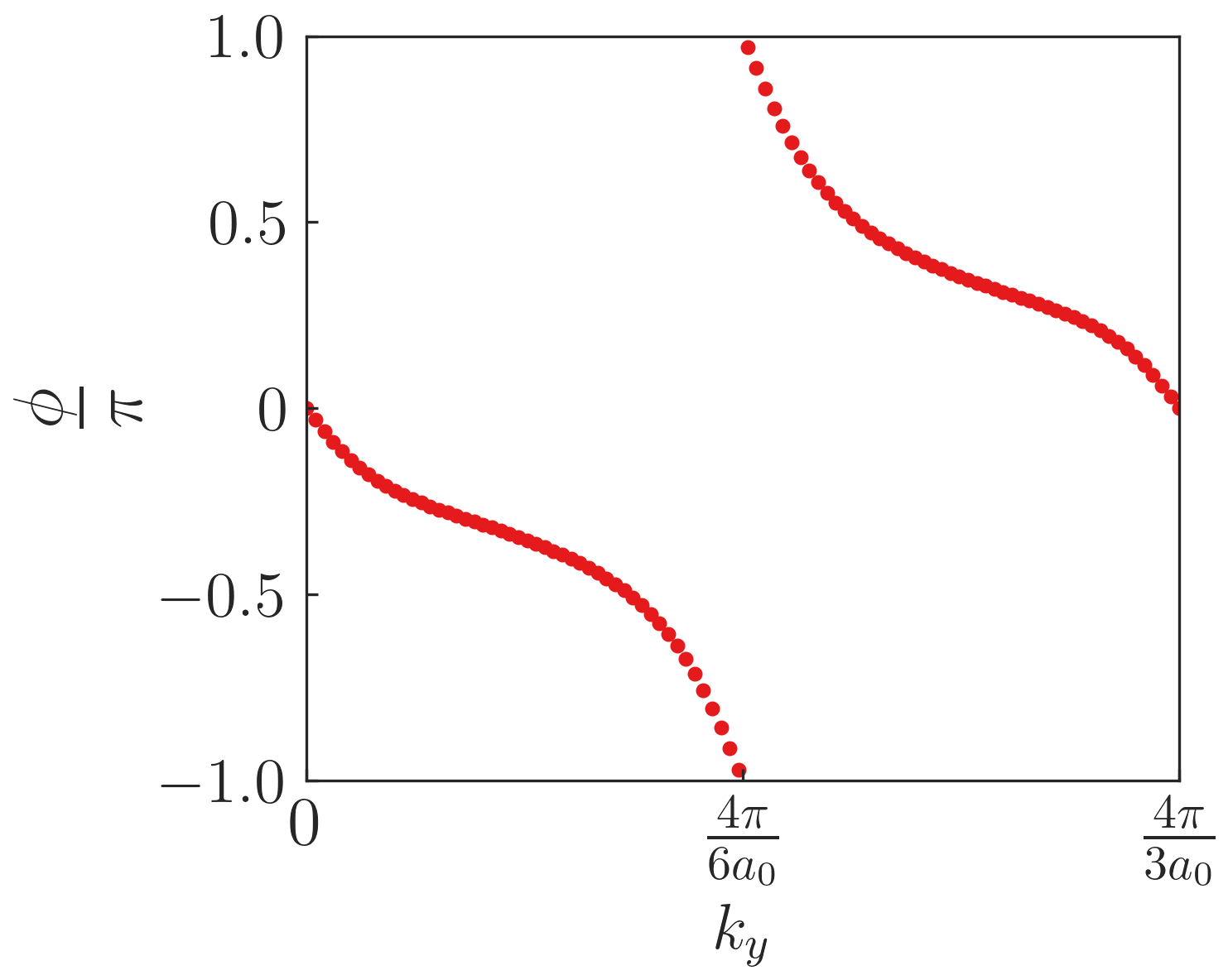}
         \caption{$\xi=1.0$}
         \label{Fig6a}
     \end{subfigure}
\begin{subfigure}[b]{0.49\columnwidth}
         \centering
         \includegraphics[width=\columnwidth]{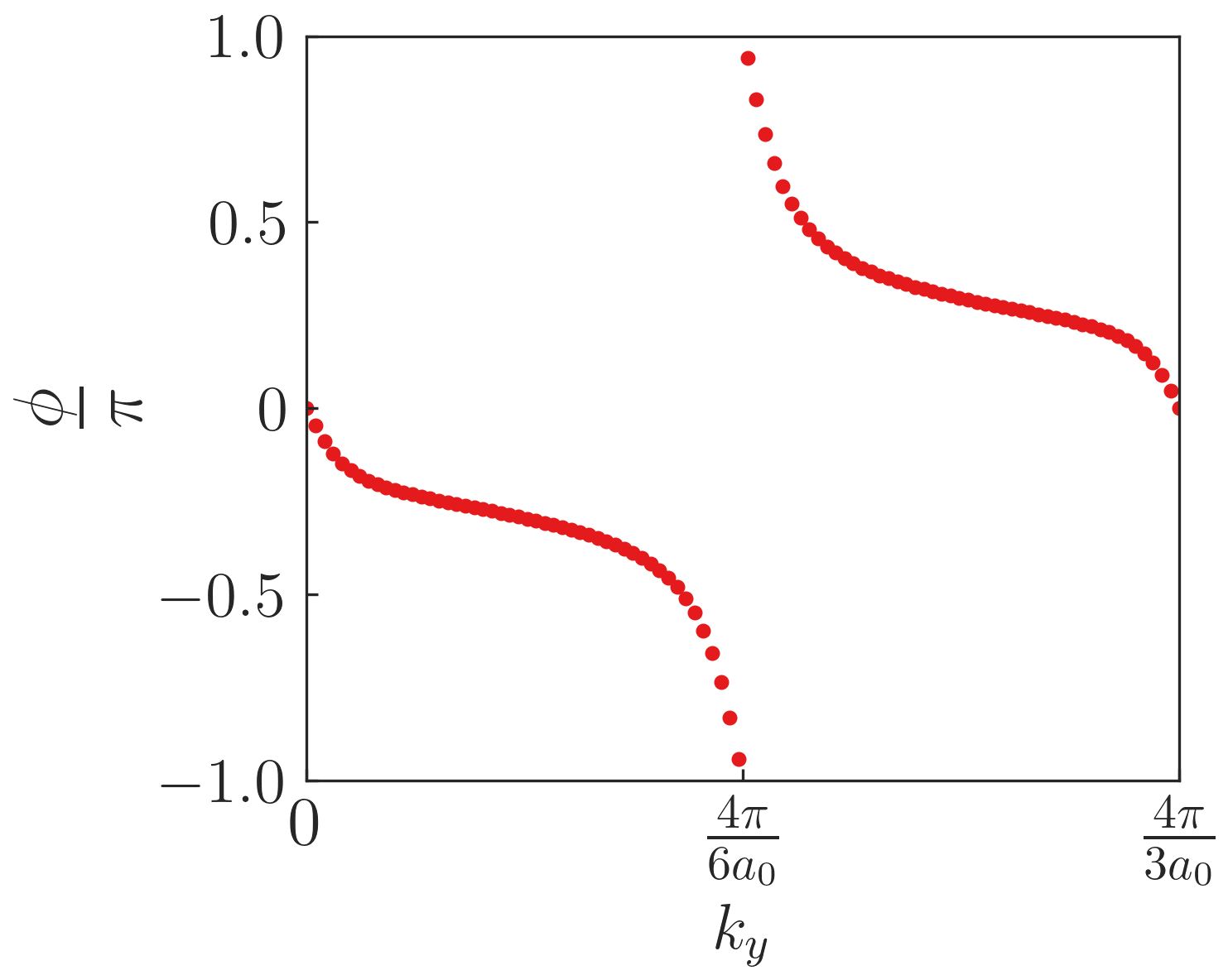}
         \caption{$\xi=1.5$}
         \label{Fig6b}
     \end{subfigure}
     \begin{subfigure}[b]{0.49\columnwidth}
         \centering
         \includegraphics[width=\columnwidth]{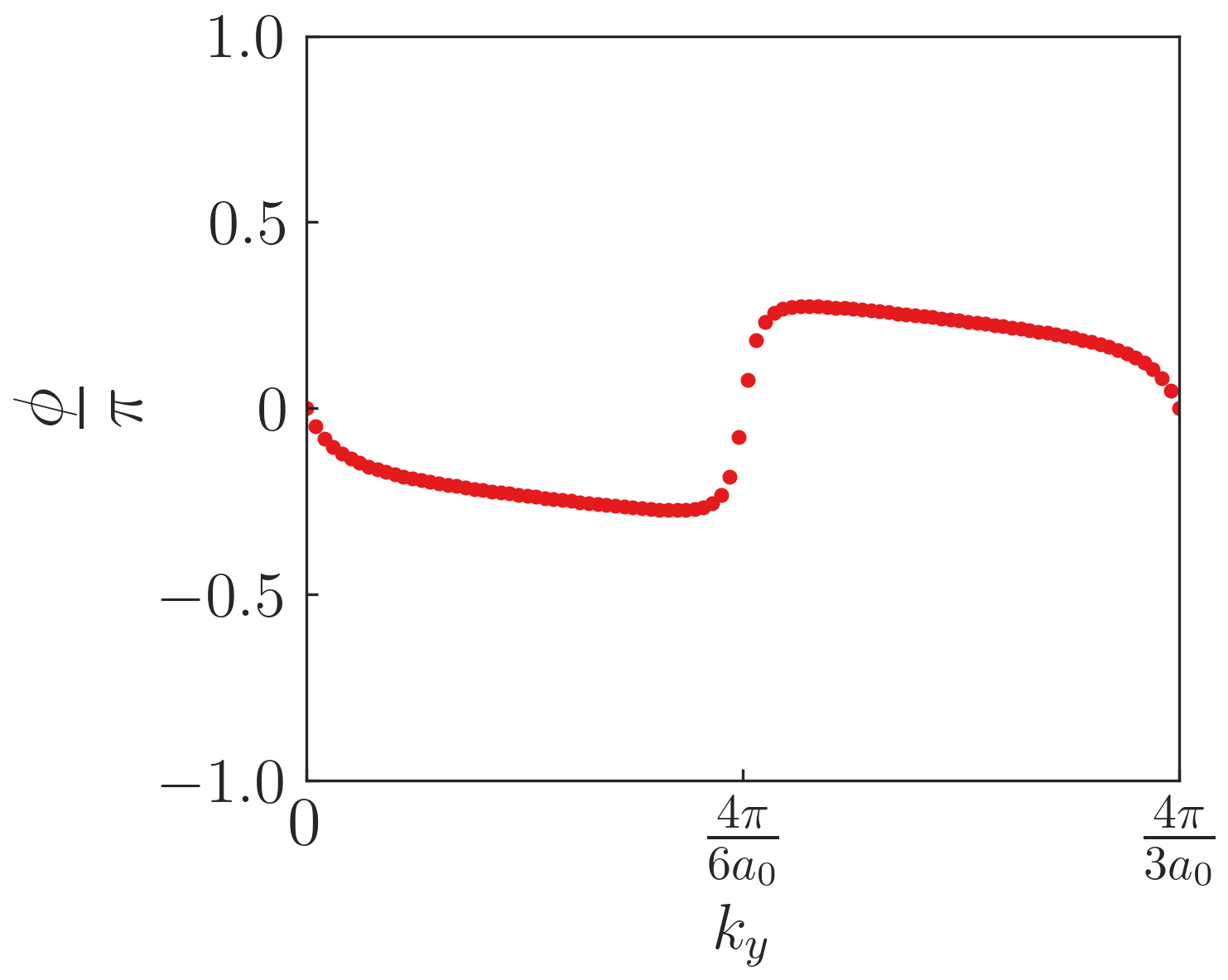}
         \caption{$\xi=2.2$}
         \label{Fig6c}
     \end{subfigure}
     \begin{subfigure}[b]{0.49\columnwidth}
         \centering
         \includegraphics[width=\columnwidth]{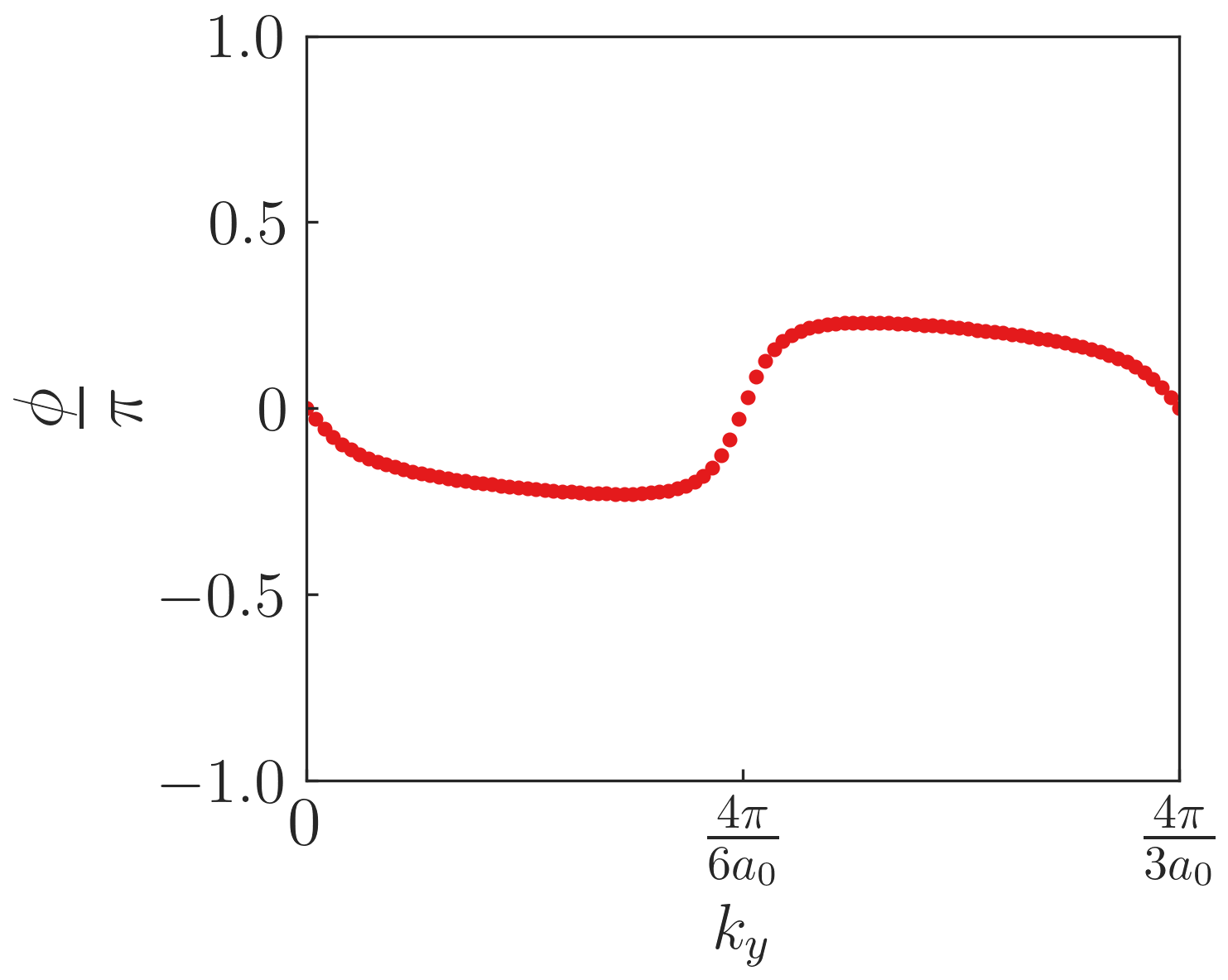}
         \caption{$\xi=2.5$}
         \label{Fig6d}
     \end{subfigure}
\caption{Wannier charge center (WCC) along the $x$-direction has been plotted as a function of momentum along the $y$-direction, namely $k_y$. (a), (b) TI phase, where a non-trivial winding of the WCC, is noted as $k_y$ traverses a closed loop in the BZ. This is equivalent to having a non-trivial Chern number and hence a non-trivial bulk. The jump in WCC indicates a shift in the average position of the electrons in a unit cell. (c), (d) The system here is no longer in the TI phase. The WCC trivially oscillates about zero as a function of $k_y$.}
 \label{Fig6}
\end{figure}

 Now, we look at the evolution of the Wannier charge center (WCC) over a closed loop in the BZ (Fig. \ref{Fig6}) and study its behaviour as a function of $\xi$. For a 2D system, this is equivalent to calculating the Berry phase along one direction (say $k_x$) and evolving it as a function of momentum in the other direction (say $k_y$) \cite{WCC_1, WCC_2}. This plot bears information about the bulk topology. In our case, we calculate the WCC along the $x$ direction and look for its evolution as a function of $k_y$. The mathematical expression for calculating the WCC (or equivalently the Berry phase) is given as \cite{PhysRevB.78.195125},
\begin{align}
    \begin{split}
        \phi_n(k_y)=\frac{1}{2\pi}\int_{0}^{2\pi}A_n(k_x,k_y)dk_x
    \end{split}
\end{align}
Here $\vec A_n(k_x, k_y)=-i\langle u_{nk}|\nabla_k|u_{nk}\rangle$, where $n$ is the band index. In the TI phase, we see the WCC winding non-trivially as $k_y$ in increased from $0$ to $2\pi$ (\ref{Fig6a}, \ref{Fig6b}). This indicates a shift in the charge center of the reference unit cell over a closed loop in $k_y$. Charge conservation in this case demands the presence of edge states that traverse the bulk energy gap and transfers an electron from the valence to the conduction band. We also plot the behaviour of the WCC in the region $\xi>2$. It can be clearly seen that the WCC shows no winding here and oscillates about zero. This hints at a topologically trivial phase.
\section{\label{sec:level3}Higher order topological phase}
\begin{figure}
    \centering
    \includegraphics[height=5cm, width=0.5\columnwidth]{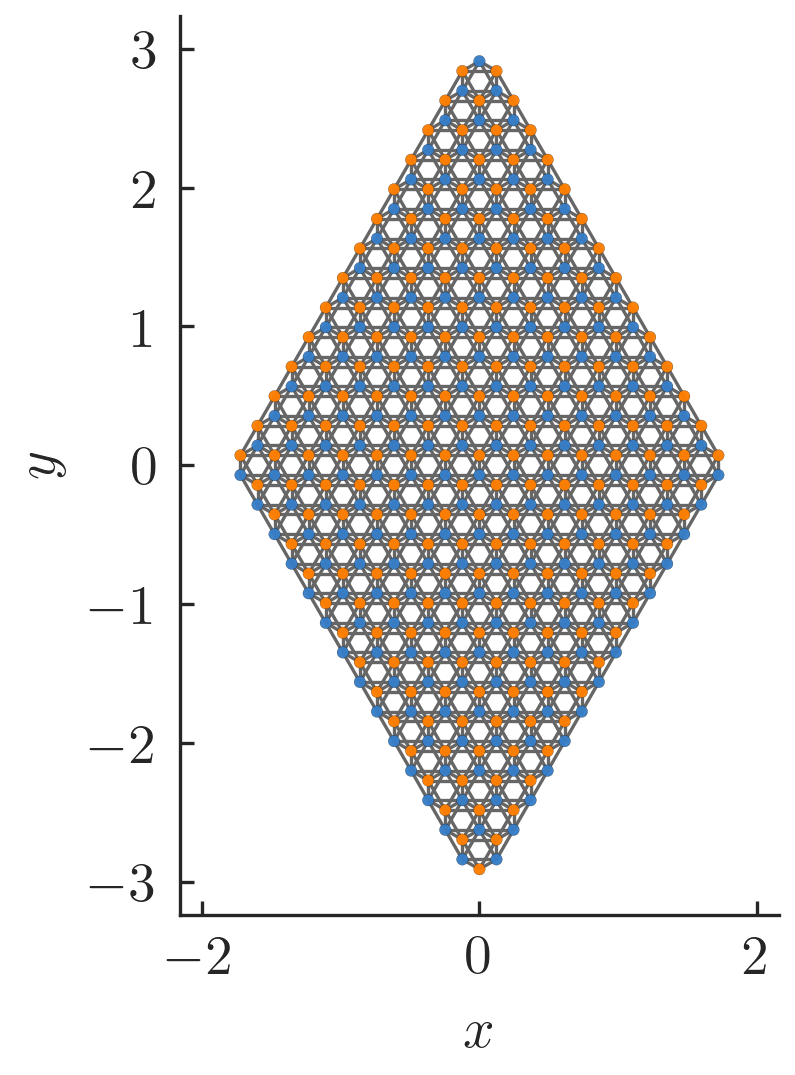}
    \caption{A rhombic supercell constructed using a honeycomb lattice which is used in our calculation for the HOTI phase.}
    \label{Fig7}
\end{figure}
\begin{figure}
\begin{subfigure}[b]{0.49\columnwidth}
         \centering
         \includegraphics[width=\columnwidth]{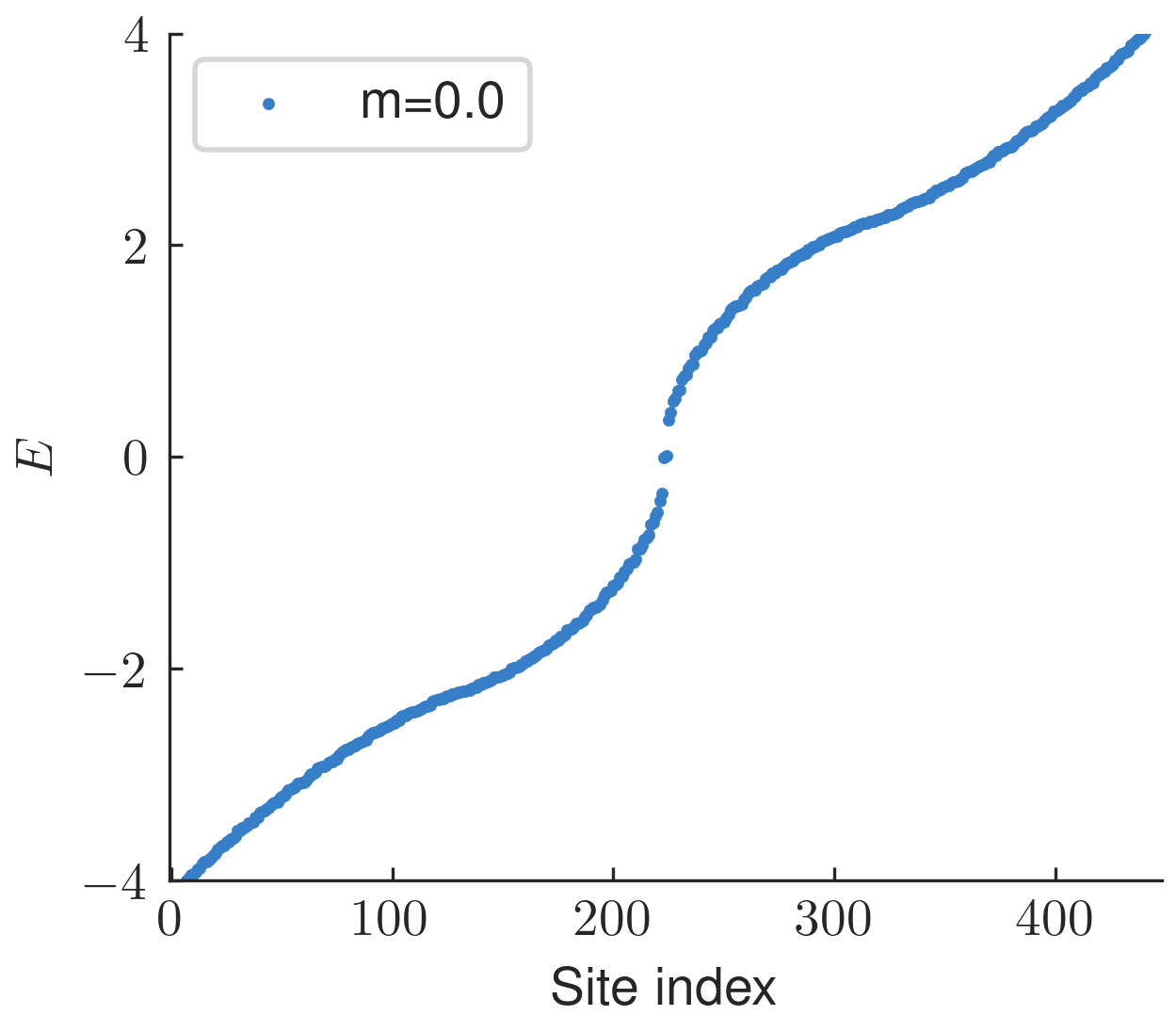}
         \caption{$\xi=2.2$}
         \label{Fig8a}
     \end{subfigure}
\begin{subfigure}[b]{0.49\columnwidth}
         \centering
         \includegraphics[width=\columnwidth]{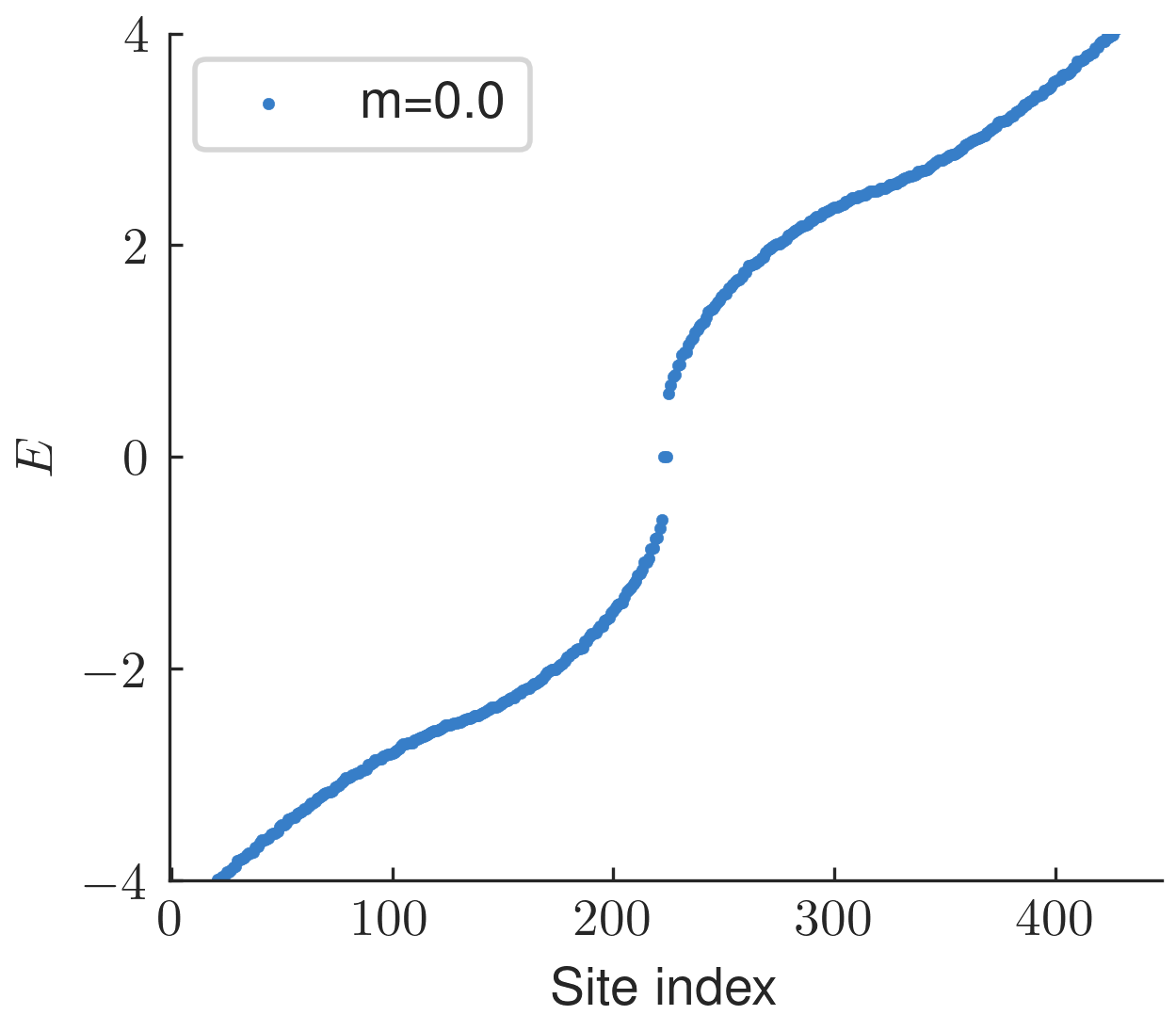}
         \caption{$\xi=2.5$}
         \label{Fig8b}
     \end{subfigure}
     \begin{subfigure}[b]{0.49\columnwidth}
         \centering
         \includegraphics[width=\columnwidth]{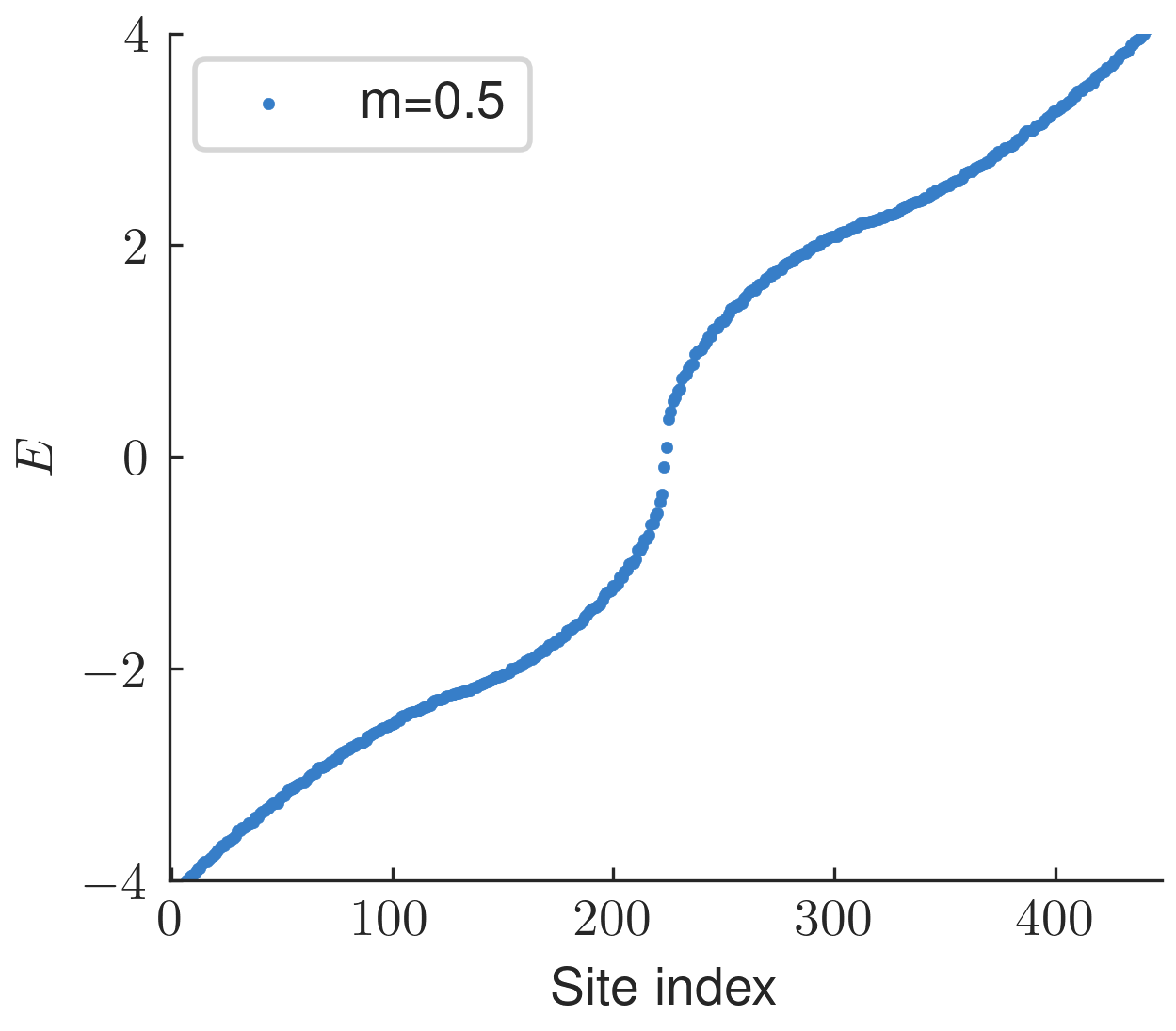}
         \caption{$\xi=2.2$}
         \label{Fig8c}
     \end{subfigure}
     \begin{subfigure}[b]{0.49\columnwidth}
         \centering
         \includegraphics[width=\columnwidth]{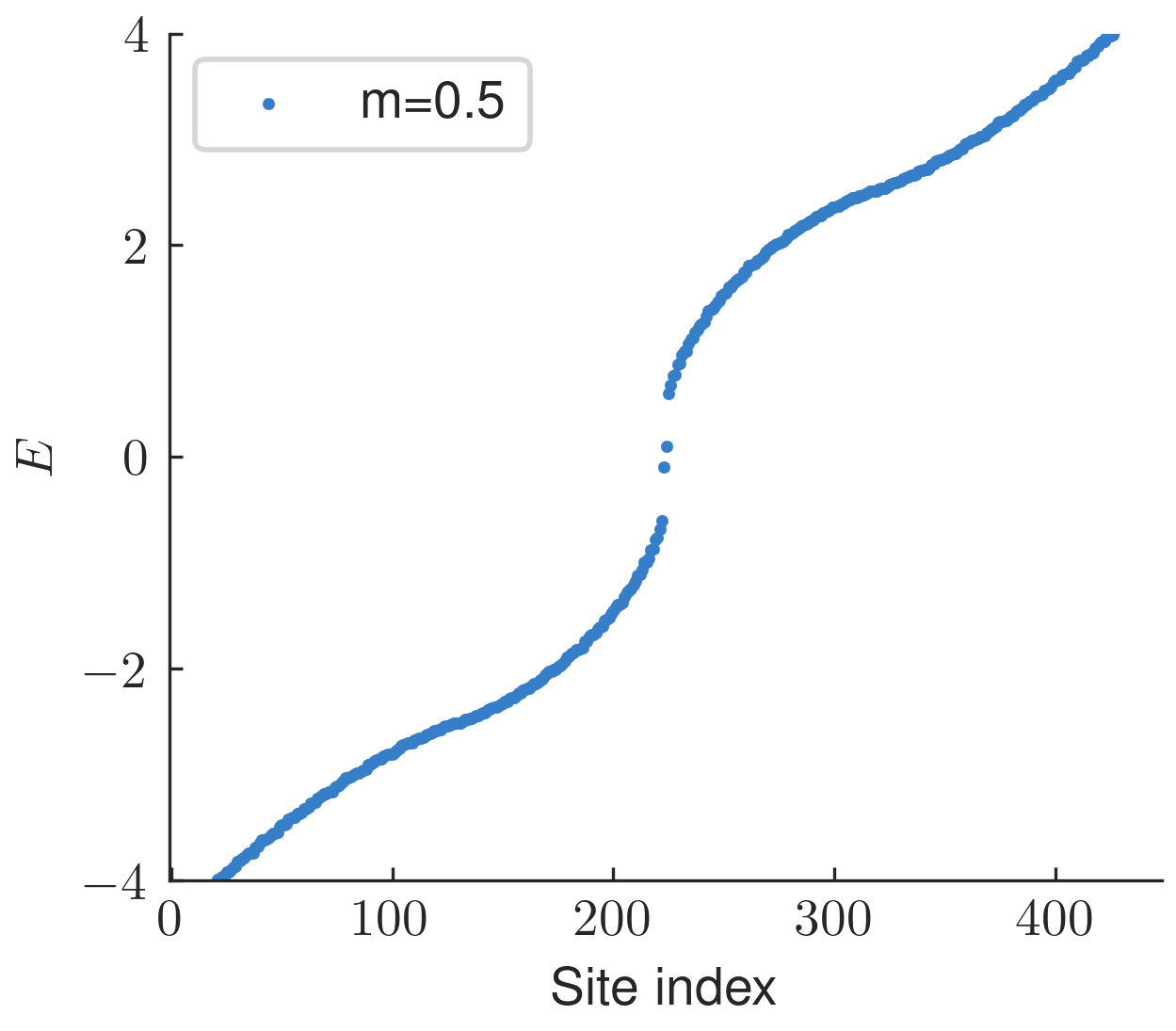}
         \caption{$\xi=2.5$}
         \label{Fig8d}
     \end{subfigure}
\caption{Real space Hamiltonian of the band deformed Haldane model has been diagonalized on the rhombic supercell. The resulting eigenvalues for (a) $\xi=2.2$ and (b) $\xi=2.5$ have been shown when the Semeneoff mass, $m=0$. The plots show the presence of zero energy eigenvalues which are distinctly separated from the bulk. It is to be noted that the system is no more in the TI phase for both the plots. In (c) and (d), the modes that were initially at zero energy, shift in the presence of a non-zero Semenoff mass.}
\label{Fig8}
\end{figure}
\begin{figure}
\begin{subfigure}[b]{0.45\columnwidth}
         \centering
         \includegraphics[width=\columnwidth]{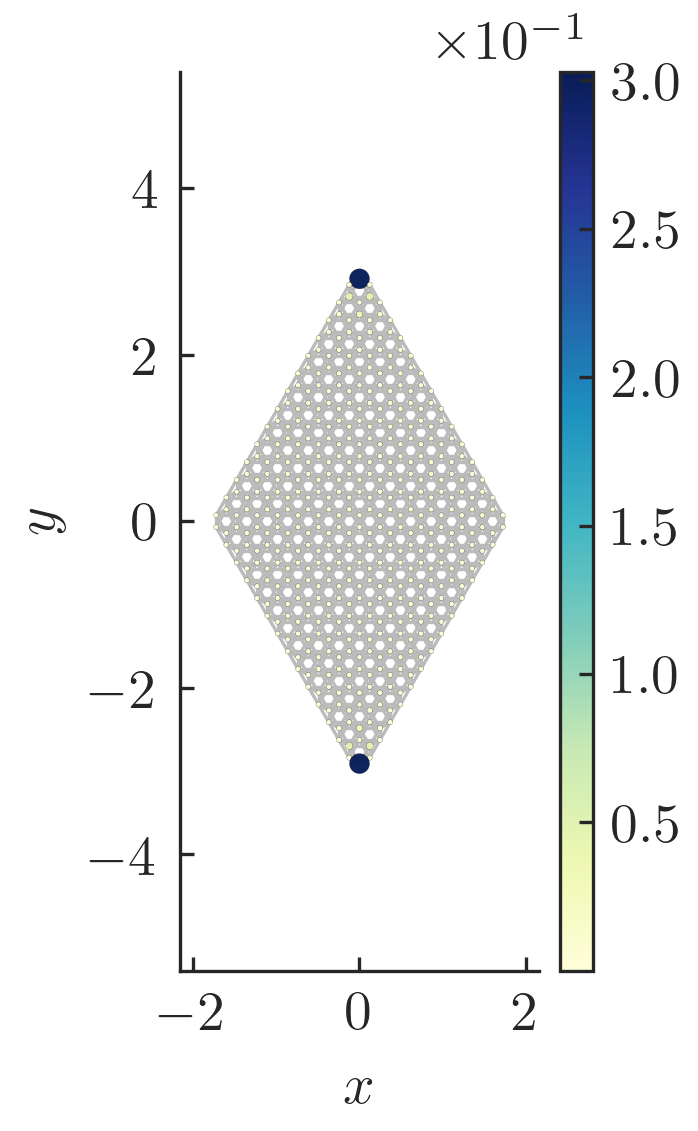}
         \caption{$\xi=2.2$}
         \label{9a}
     \end{subfigure}
\begin{subfigure}[b]{0.45\columnwidth}
         \centering
         \includegraphics[width=\columnwidth]{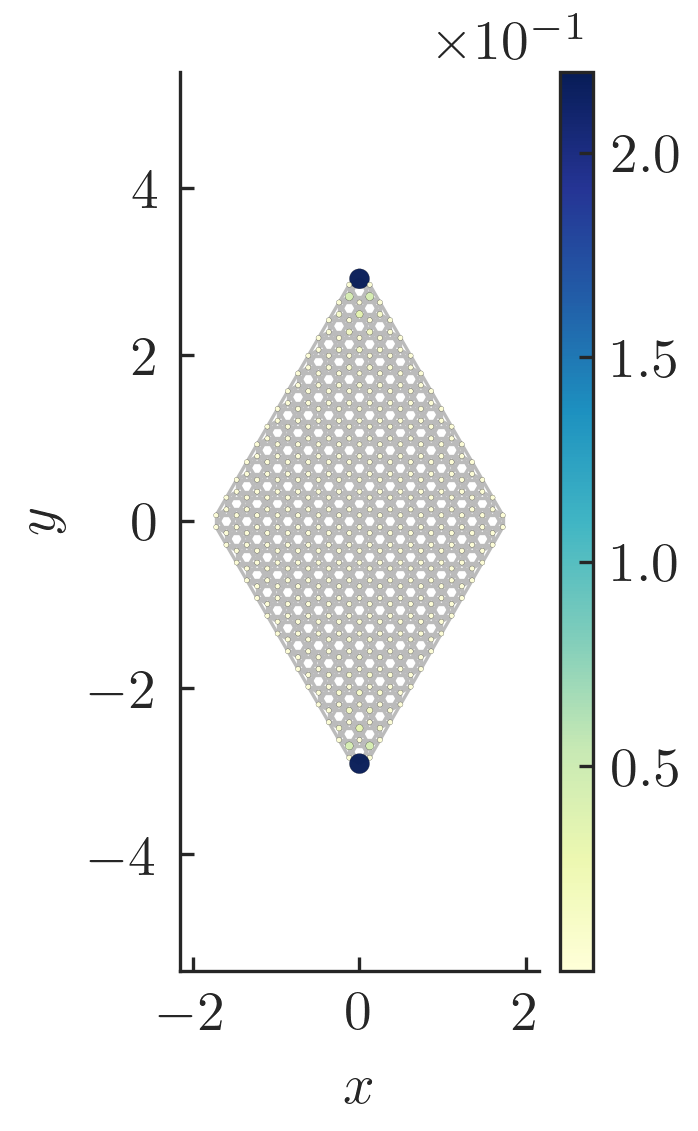}
         \caption{$\xi=2.5$}
         \label{9b}
     \end{subfigure}
\caption{The probability density plots for the states at zero energy corresponding to the parameter values (a) $\xi=2.2$ and (b) $\xi=2.5$. The system is in the HOTI phase and demonstrates the presence of corner modes shown by deep blue colour.}
\label{Fig9}
\end{figure}

We now focus on the region where $\xi>2$. The Chern number and the WCC both hint at this region being topologically trivial. We, however take a suitably formed rhombic supercell (Fig. \ref{Fig7}) and diagonalize the real space Hamiltonian for this structure. The energy spectra of the real space Hamiltonian, for $\xi<2$ does not show any specific feature with regard to higher order topology. However beyond the TI phase ($\xi\geq 2$), we see the presence of modes pinned at zero energy and distinctly separated from the bulk (Fig \ref{Fig8}). We further plot the probability density of these states which are found to be localized at two corners of the rhombus (Fig. \ref{Fig9}). We had kept the value of the Semenoff mass $m$ fixed at $0$ which keeps the inversion symmetry in the system intact. On increasing the Semenoff mass, inversion symmetry in the system gets broken. The degeneracy of the states is lifted and the states drift from zero energy. They are, however no longer cleanly separated from the bulk.\par
To characterize the corner states of the rhombic supercell, we calculate the bulk polarisation which is equivalent to the position of charge in a unit cell. In a crystal lattice it is difficult to unambiguously define a well defined unit cell that hosts a periodic distribution of charge. Such a definition is however important for the evaluation of polarisation under the Clausius-Mosotti picture \cite{Resta2007}. Hence, in bulk solids, we rely on the Wannier functions for the calculation of bulk polarisation. Wannier functions are localized wave functions formed by suitable superposition of the delocalized Bloch wave functions. The expectation value of the position operator with respect to these Wannier functions is known as the Wannier charge center and is used to define polarisation for a crystalline system. WCC, and hence the polarisation in a direction $\alpha$ is given as \cite{Pyrochlore}
\begin{align}
\begin{split}
 p_\alpha={\bar r_\alpha} = \langle w_n|r_\alpha|w_n\rangle=\frac{i}{S}\int_{BZ} d^{d}k \langle u_{nk}|\frac{\partial}{\partial k_{r_\alpha}}|u_{nk}\rangle
\end{split}
\end{align}
\begin{figure}
\begin{subfigure}{0.49\columnwidth}
         \centering
         \includegraphics[width=\columnwidth]{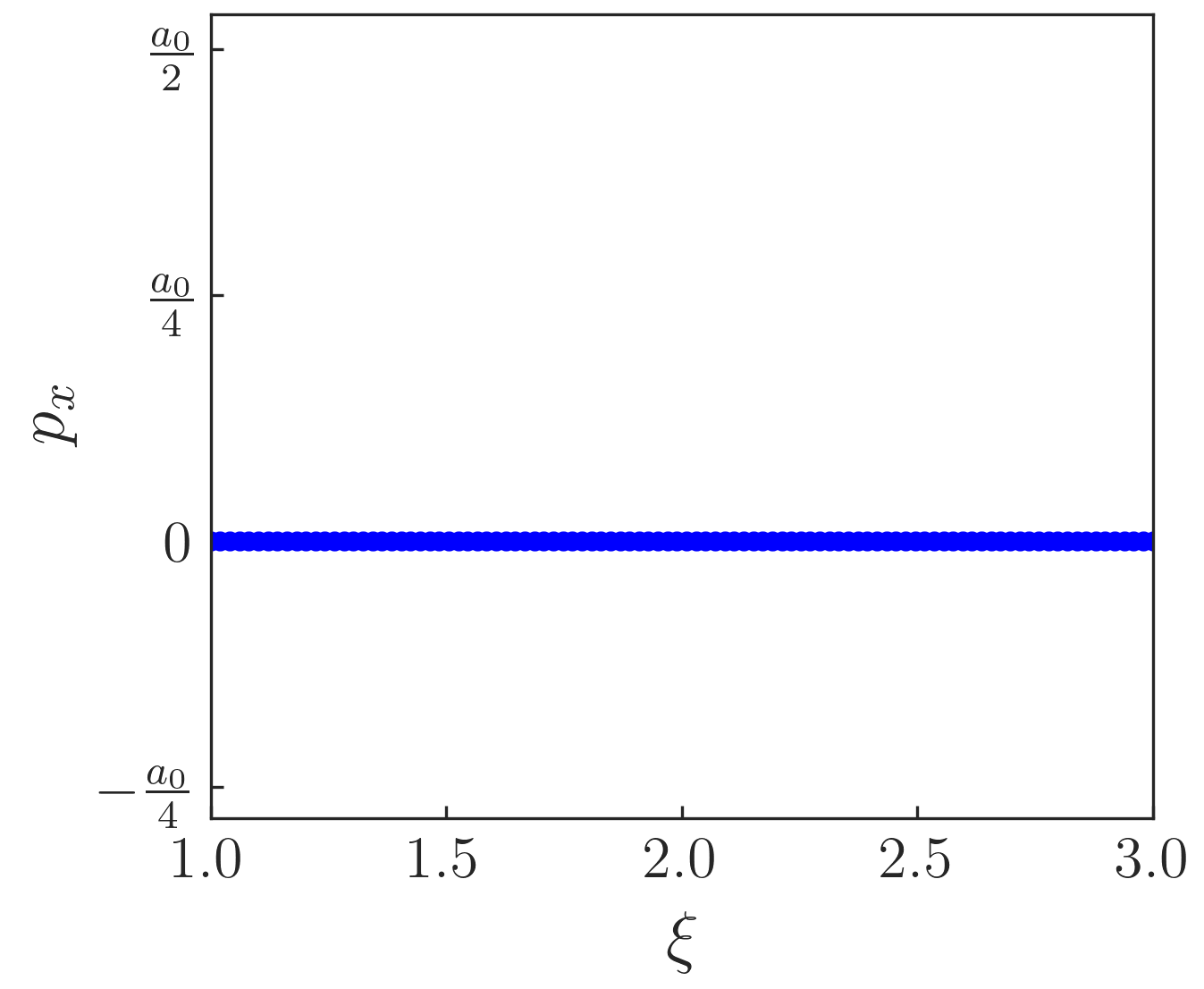}
         \caption{}
         \label{Fig10a}
     \end{subfigure}
\begin{subfigure}{0.49\columnwidth}
         \centering
         \includegraphics[width=\columnwidth]{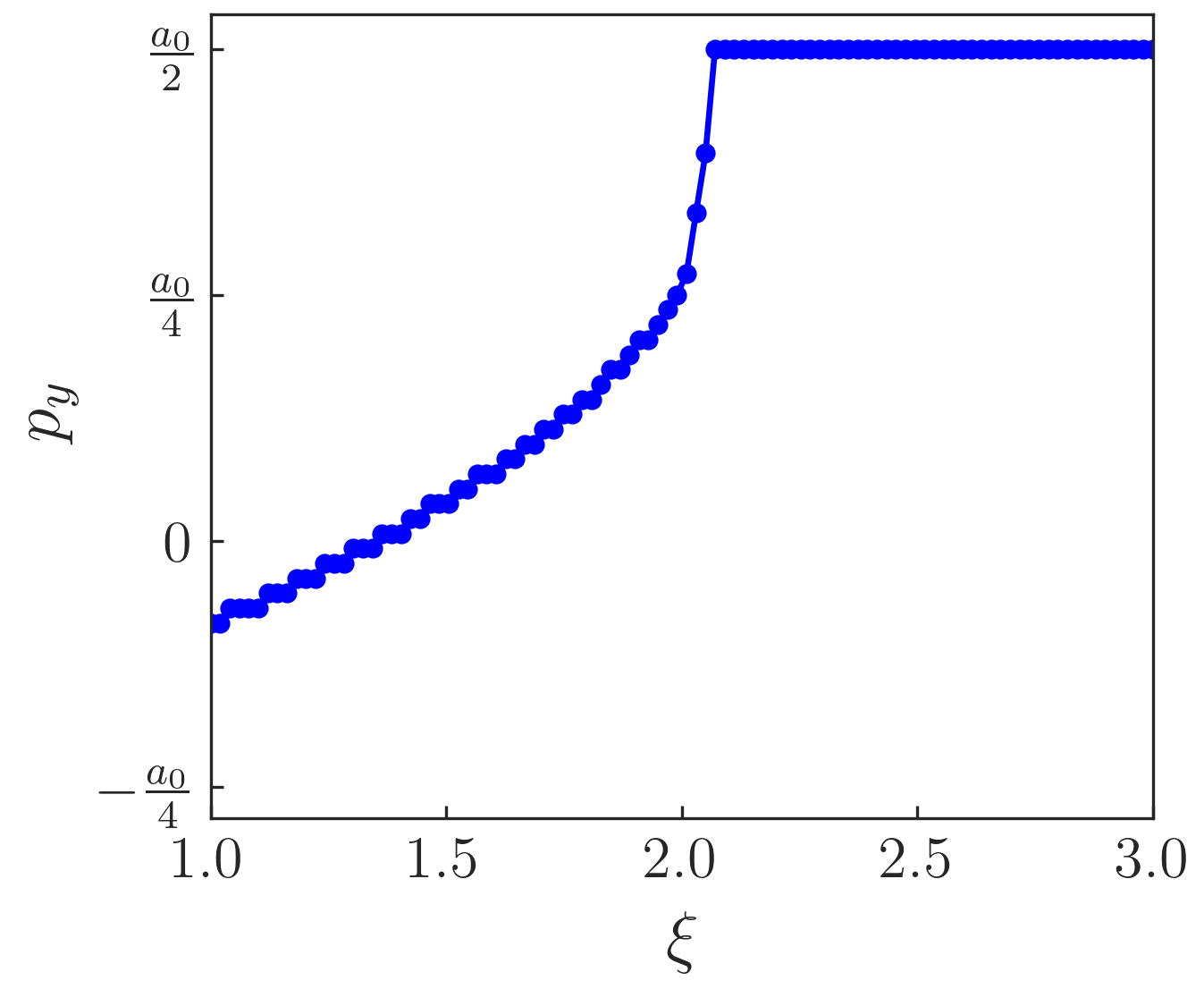}
         \caption{}
         \label{Fig10b}
     \end{subfigure}
\caption{The polarisation, $p_x$, $p_y$ as a function of $\xi$ has been shown.(a) $p_x$ vanishes uniformly both in the TI and HOTI phase. (b) $p_y$ is quantized in the HOTI phase at $\frac{a_0}{2}$ which is the center of the strong bond.}
\label{Fig10}
\end{figure}
where $|w_n\rangle$ is the Wannier function corresponding to the $n^{th}$ band. Here we have taken the electronic charge $e$ to be $1$. $S$ is the area of the 2D BZ of the honeycomb lattice which in this case is equal to $\frac{8\pi^2}{3\sqrt{3}a_0^2}$. It can be seen that the polarisation is directly related to the Berry connection of the system. Owing to the gauge dependence of the Berry connection, the polarisation is defined modulo a lattice vector. We calculate $p_{\alpha=x,y}$ as a function of $\xi$ for our deformed Haldane model (Fig. \ref{Fig10}). $p_y$ is found to be perfectly quantized at $\frac{a_0}{2}$ for $\xi>2$. For $\xi<2$ (TI phase), the value of $p_y$ is not quantized. In case of the polarisation along the $x$ direction, we find that it is uniformly zero both for the TI and HOTI phase. The quantized value of $p_y$ at $\frac{a_0}{2}$ indicates a mismatch between the positions of the WCC and the lattice site. This indicates a non-trivial bulk topology. The quantization of the polarisation $p_y$ is due to the presence of inversion symmetry in the system. Inversion symmetry (which is only present when $m=0$) allows an interchange of the A and B sublattices, while keeping the center of the strong bond invariant. This causes the center of charge to be localized at the center of the strong bond and results in $p_y$ to be quantized.
{\section{\label{sec:level4}Conclusion}}
In this work, we have studied a band deformed Haldane model by smoothly tuning one of its nearest neighbour hopping amplitudes. This band deformation breaks the $C_3$ symmetry of the original Haldane model and causes the Dirac nodes at the \textbf{K} and \textbf{K$'$} points in the BZ to shift and move towards an intermediate \textbf{M} point as $t_1$ (the nearest neighbour hopping amplitude along $\hat \delta_1$) approaches $2t$ (where $t$ is the nearest neighbour hopping amplitude along $\hat \delta_2$ and $\hat \delta_3$). We show that this causes a shift in the concentration of Berry curvature in the BZ as well.  We calculate Chern number and plot it as a function of $\xi$($=\frac{t_1}{t}$) to show that the Chern number has a value of $1$ for $\xi<2$ while it becomes zero beyond that. We establish that the region $\xi>2$ is a higher order topological phase, as contrary to earlier works which reported this phase as trivial owing to the absence of edge states in the system. We calculate bulk polarisation ($p_{\alpha=x,y}$) for this model, which serves as a topological invariant. $p_y$ remains constant at $\frac{a_0}{2}$ while $p_x$ remains zero for the entire HOTI phase. This quantization is a consequence of the inversion symmetry which is intact in the system as long as the semenoff mass remains zero.
\bibliographystyle{ieeetr}
\bibliography{ref.bib}
\end{document}